\documentclass[sigconf]{acmart}
\pdfoutput=1

\copyrightyear{2024}
\acmYear{2024}
\setcopyright{rightsretained}
\acmConference[SPAA '24]{Proceedings of the 36th ACM Symposium on Parallelism in Algorithms and Architectures}{June 17--21, 2024}{Nantes, France}
\acmBooktitle{Proceedings of the 36th ACM Symposium on Parallelism in Algorithms and Architectures (SPAA '24), June 17--21, 2024, Nantes, France}\acmDOI{10.1145/3626183.3659985}
\acmISBN{979-8-4007-0416-1/24/06}

\makeatletter
\gdef\@copyrightpermission{
\begin{minipage}{0.3\columnwidth}
 \href{https://creativecommons.org/licenses/by/4.0/}{\includegraphics[width=0.90\textwidth]{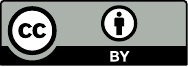}}
\end{minipage}\hfill
\begin{minipage}{0.7\columnwidth}
 \href{https://creativecommons.org/licenses/by/4.0/}{This work is licensed under a Creative Commons Attribution International 4.0 License.}
\end{minipage}
\vspace{5pt}
}
\makeatother

\newif\iftr     
\newif\ifall    
\newif\ifconf   
\newif\ifsq     
\newif\ifnonb   
\newif\iftodos  

\newif\ifsqCAP
\newif\ifsqVS
\newif\ifsqEN
\newif\ifsqTIT

\DeclareMathOperator{\nnz}{nnz}
\DeclareUnicodeCharacter{202F}{\,}
\newcommand{\ignore}[1]{}
\sqtrue
\sqCAPtrue
\sqENtrue
\sqVStrue
\sqTITtrue

\newboolean{showStriked}
\setboolean{showStriked}{false}

\newcommand{\conditionalPara}[1]{
    \ifthenelse{\boolean{showStriked}}{\sout{#1}}{}
}

\usepackage{etex}

\usepackage{epstopdf}
\usepackage{placeins}
\usepackage{adjustbox}

\usepackage{mathtools}
\usepackage{afterpage}
\usepackage{tensor}

\usepackage{graphicx}
\usepackage{float}
\usepackage{dblfloatfix}
\usepackage{multirow}
\usepackage{rotating}
\usepackage[switch]{lineno} 
\usepackage{makecell}
\usepackage{tabulary}
\usepackage{parcolumns}
\usepackage{tikz}
\usepackage{bm}
\usepackage{caption}
\usepackage{subcaption}
\usepackage{xr}
\usepackage{cases}
\usepackage{xpatch}
\expandafter\xpatchcmd
\csname pgfk@/tikz/every picture/.@cmd\endcsname
{\thepage}{\arabic{page}}{}{}

\tikzstyle{comment} = [draw, fill=blue!70, text=white, text width=3cm, minimum height=1cm, rounded corners, align=left, font=\scriptsize]
\tikzstyle{background_alg} = [draw, fill=blue!20, opacity=0.4, inner sep=4pt, rounded corners=2pt]

\usetikzlibrary{shapes}
\usetikzlibrary{plotmarks}
\usetikzlibrary{calc, fit}

\usepackage{enumitem}

\usepackage{amsthm}
\usepackage{mathtools,mathrsfs}

\newtheorem{theorem}{Theorem}[section]

\newtheorem{definition}[theorem]{Definition}

\usepackage{soul}
\usepackage{fontawesome}
\usepackage{pifont}
\usepackage{textcomp}
\usepackage{booktabs}
\usepackage{url}
\usepackage{pbox}
\usepackage[normalem]{ulem}
\usepackage[10pt]{moresize}

\usepackage{cleveref}
\usepackage[utf8]{inputenc}
\usepackage{algorithm}
\usepackage{algcompatible}
\usepackage[noend]{algpseudocode}

\usepackage[customcolors]{hf-tikz}
\hfsetbordercolor{white}
\hfsetfillcolor{vlgray}

\definecolor{vlgray}{rgb}{0.77 0.77 0.77}
\definecolor{ablack}{rgb}{0.2 0.2 0.2}
\definecolor{vllgray}{rgb}{0.94 0.94 0.94}
\definecolor{xblue}{rgb}{0.4 0.4 0.99}
\definecolor{bblue}{rgb}{0.5 0.7 0.7}
\definecolor{lblack}{rgb}{0.5 0.5 0.5}

\usepackage[indLines=true,noEnd=true,commentColor=lblue,endLComment=]{algpseudocodex}
\algrenewcommand{\ALG@beginalgorithmic}{\ttfamily}
\algnewcommand{\LineComment}[1]{\State \(\triangleright\) #1}

\usetikzlibrary{fit,tikzmark}

\newcounter{tmkcount}
\tikzset{%
    tikzmark suffix={-\thetmkcount},%
    defaultCodeBox/.style={draw=red}%
}

\newcommand{\drawCodeBox}[4]{%
    \begin{tikzpicture}[remember picture,overlay]
        \coordinate (start) at ([yshift=1.4ex]pic cs:#2);
        \coordinate (middle) at (pic cs:#3);
        \coordinate (end) at ([yshift=-0.2ex]pic cs:#4);
        \node[inner sep=2pt,#1,fit=(start) (middle) (end)] {};
    \end{tikzpicture}%
}

\newcommand{\BeginBoxx}[1]{%
    \drawCodeBox{#1}{beginCB}{middleCB}{endCB}%
    \tikzmark{beginCB}\tikzmark{middleCB}%
}

\newcommand*\circled[1]{%
  \protect\tikz[baseline=(C.base)]\protect\node[draw,circle,inner sep=0.5pt](C) {#1};\!
}

\algdef{S}[WHILE]{BoxedWhile}[2][defaultCodeBox]{\BeginBoxx{#1}\algorithmicwhile\ #2\ \algorithmicdo}%
\algdef{S}[FOR]{BoxedFor}[2][defaultCodeBox]{\BeginBoxx{#1}\algorithmicfor\ #2\ \algorithmicdo}%
\algdef{S}[FOR]{BoxedForAll}[2][defaultCodeBox]{\BeginBoxx{#1}\algorithmicforall\ #2\ \algorithmicdo}%
\algdef{S}[LOOP]{BoxedLoop}[1][defaultCodeBox]{\BeginBoxx{#1}\algorithmicloop}%
\algdef{S}[REPEAT]{BoxedRepeat}[1][defaultCodeBox]{\BeginBoxx{#1}\algorithmicrepeat}%
\algdef{S}[IF]{BoxedIf}[2][defaultCodeBox]{\BeginBoxx{#1}\algorithmicif\ #2\ \algorithmicthen}%
\algdef{C}[IF]{IF}{BoxedElsIf}[2][defaultCodeBox]{\BeginBoxx{#1}\algorithmicelse\ \algorithmicif\ #2\ \algorithmicthen}%
\algdef{Ce}[ELSE]{IF}{BoxedElse}{EndIf}[1][defaultCodeBox]{\BeginBoxx{#1}\algorithmicelse}%

\ifsqCAP
\else
\fi

\newcommand{\enlargeSQ}[1]{\ifsqEN\enlargethispage{\baselineskip}\fi}
\ifsqTIT
\fi

\usepackage{xcolor}
\definecolor{darkgrey}{RGB}{70,70,70}
\definecolor{lightgrey}{RGB}{200,200,200}
\definecolor{lyellow}{RGB}{255,255,100}
\definecolor{llyellow}{RGB}{250,250,180}
\definecolor{lgreen}{RGB}{40,120,40}
\definecolor{lblue}{RGB}{0,76,153}
\definecolor{lred}{RGB}{51,102,0}
\definecolor{lblack}{RGB}{64,64,64}
\definecolor{lgren}{rgb}{165,0,0}

\usepackage{colortbl}

\usepackage{listings}

\ifsq
\else
\fi

\definecolor{hlL}{rgb}{0.8 0.8 0.99}
\definecolor{hlLB}{rgb}{0.63 0.79 0.95}

\newcounter{highlight}

\newcounter{hlLR}

\newcounter{hlLIR}

\newcounter{hlLIIR}

\newcounter{hlLRX}

\newcounter{gbox}

\newcounter{vgbox}

\newcounter{ggbox}

\newcounter{gboxblue}

\newcounter{hlLRB}

\newcounter{Ahighlight}

\usepackage{textcomp}
\DeclareMathSymbol{\mathdblquotechar}{\mathalpha}{letters}{`"}

\newcommand{\mathdblquote}{\mathtt{\mathdblquotechar}}
\begingroup\lccode`~=`"\lowercase{\endgroup
  \let~\mathdblquote
}
\AtBeginDocument{\mathcode`"="8000 }

\makeatletter
\AtBeginDocument{\let\c@lstlisting\c@figure}
\makeatother

\usepackage{scalerel,stackengine}
\stackMath
\newcommand\rwh[1]{%
\savestack{\tmpbox}{\stretchto{%
  \scaleto{%
        \scalerel*[\widthof{\ensuremath{#1}}]{\kern-.6pt\bigwedge\kern-.6pt}%
                  {\rule[-\textheight/2]{1ex}{\textheight}}
                              }{\textheight}%
}{0.5ex}}%
\stackon[1pt]{#1}{\tmpbox}%
}

\usepackage[hang,flushmargin]{footmisc}

\renewcommand{\epsilon}{\ensuremath\varepsilon}

\renewcommand{\phi}{\ensuremath{\varphi}}

\usepackage{comment}
\usepackage[mathscr]{eucal}
\usepackage{balance}
\newcommand{\tsr}[1]{\mathcal{#1}}

\conftrue
\allfalse
\trfalse
\todosfalse

\newif\iftrLOW
\trLOWfalse

\newif\iftrNEW
\trNEWfalse

\newif\iftrDYN
\trDYNfalse

\usepackage{colortbl}

\newcommand{\marginparN}[1]{\marginpar{}}
\newcommand{\marginparX}[1]{}

\begin{document}
\setlength{\abovedisplayskip}{5pt}
\setlength{\abovedisplayshortskip}{5pt}
\setlength{\belowdisplayskip}{5pt}
\setlength{\belowdisplayshortskip}{5pt}
\title[Minimum Cost Loop Nests for Contraction of a Sparse Tensor with a Tensor Network]{Minimum Cost Loop Nests for Contraction of a Sparse Tensor with a Tensor Network}


\author{Raghavendra Kanakagiri}
\affiliation{
  \department{Department of Computer Science}              
  \institution{University of Illinois Urbana-Champaign}            
  \city{Urbana}
  \state{IL}
  \country{USA}                    
  \institution{}
  \institution{Indian Institute of Technology Tirupati}            
  \city{Tirupati}
  \state{AP}
  \country{India}                    
}
\email{raghavendra@iittp.ac.in}          

\author{Edgar Solomonik}
\affiliation{
  \department{Department of Computer Science}              
  \institution{University of Illinois Urbana-Champaign}            
  \city{Urbana}
  \state{IL}
  \country{USA}                    
}
\email{solomon2@illinois.edu}         

\begin{abstract}
Sparse tensor decomposition and completion are common in numerous applications, ranging from machine learning to computational quantum chemistry.
Typically, the main bottleneck in optimization of these models are contractions of a single large sparse tensor with a network of several dense matrices or tensors (SpTTN). 
Prior works on high-performance tensor decomposition and completion have focused on performance and scalability optimizations for specific SpTTN kernels.
We present algorithms and a runtime system for identifying and executing the most efficient loop nest for any SpTTN kernel.
We consider both enumeration of such loop nests for autotuning and efficient algorithms for finding the lowest cost loop nest for simpler metrics, such as buffer size or cache miss models.
Our runtime system identifies the best choice of loop nest without user guidance, and also provides a distributed-memory parallelization of SpTTN kernels.
We evaluate our framework using both real-world and synthetic tensors. Our results demonstrate that our approach outperforms available generalized state-of-the-art libraries and matches the performance of specialized codes.
\end{abstract}



\maketitle

\section{Introduction}
\sloppy
Tensors provide a mathematical representation for multi-dimensional arrays, enabling basic operations such as contraction (composition) and decomposition of tensors.
Tensor contraction and decomposition are used in many methods for modeling quantum systems~\cite{markov2008simulating,khoromskaia2018tensor,orus2014advances,hirata2003tensor} and to construct models of data in machine learning~\cite{perros2015sparse, hutter2022high, kreimer2013tensor, cao2014robust}, as well as many other applications.
Tensor sparsity arises as a result of numerical zeros in the tensors (e.g., due to a negligible interaction as a result of physical distance between particles), or due to not all tensor entries being observed (for example, in tensor completion~\cite{tcompletion}).
Contraction of sparse tensors poses a computational challenge, due to the plethora of possible contractions and decompositions for tensors with 3 dimensions or more.

\sloppy
Acceleration of sparse tensor algebra has been pursued via runtime libraries like Cyclops Tensor Framework (CTF)~\cite{ctf}, Tensor Contraction Library (TCL)~\cite{TCL}, TiledArray~\cite{tiledarray,tiledarray_software}, Fastor~\cite{Fastor}, libtensor~\cite{libtensor}, ITensor~\cite{itensor}, Local Integrated Tensor Framework (LITF)~\cite{kats2013sparse}; code generation frameworks like TACO~\cite{TACO}, COMET~\cite{comet}, Tensor Contraction Engine (TCE)~\cite{TCE} and also specialized hardware like ExTensor~\cite{extensor}, Tensaurus~\cite{tensaurus} and Hasco~\cite{hasco}. 
These prior works have focused on enabling generalized contraction of any number of tensors. Additionally, efficient contraction of two sparse or dense tensors has also received attention, 
SpMM~\cite{comm_spmm}, SpTTM~\cite{spttm_jiajiali}, SpTV~\cite{sptv}, GEMM-like Tensor-Tensor multiplication~\cite{TCL} and contraction of two sparse tensors (SpTC)~\cite{Sparta}.
However, in the context of tensor decomposition and completion, all of the most important kernels involve contraction of a single sparse tensor (the input dataset) and many smaller dense tensors (representing the decomposition).
Such kernels have a single fixed sparsity pattern, unlike contractions such as sparse matrix multiplication, for which the cost and output sparsity is data dependent (dependent on the position of nonzeros).
We leverage the data-independent nature of sparse tensor times tensor network (SpTTN) kernels (defined generally in Section~\ref{sec:spttn_kernels}), to automatically and efficiently find minimum cost implementations.



Prior works with a focus on high-performance tensor decomposition and completion have introduced efficient and parallel implementations for many SpTTN kernels~\cite{SPLATT, HyperTensor, AdaTM, MTTKRP_Emerging_Arch, MTTKRP_Blocking, parti}.
Most of these works focus specifically on one or two kernels needed for a particular tensor decomposition, e.g., the matricized Khatri-Rao product (MTTKRP) for CP decomposition~\cite{DFacTo,GigaTensor,CSTF} or the tensor times matrix chain (TTMc) kernel for Tucker~\cite{accelerate-tucker,ptucker}).
Even for a single decomposition, different algorithms often rely on different SpTTN kernels~\cite{SINGH2022269}.
By developing algorithms and libraries for arbitrary SpTTN kernels, we provide functionality for contraction arising (e.g., as a result of a gradient calculation, described in Section~\ref{sec:spttn_kernels}) from any decomposition/network consisting of dense tensors.

The main challenge in implementation of an SpTTN kernel is finding the most efficient loop nest.
In line with prior work~\cite{TACO}, we represent such loop nest as a tree, in which each vertex is a loop and its descendants are the loops contained within it.
In Section~\ref{sec:algorithm}, we show how to enumerate all loop nests (assuming fusion is done wherever possible) for a given SpTTN.
Since each loop order for any pair of contracted tensors yields a distinct loop nest, the size of this space grows factorially in the loop nest depth $m$ and exponentially in the number of tensors $N$.
We provide a dynamic programming algorithm to find a cost-optimal loop nest with substantially lower cost, namely $O(N^32^mm)$ instead of $O((m!)^N)$.
We state the algorithm for a general cost function that can be decomposed according to the loop nest tree structure, then provide specific cost functions to minimize buffer size and cache misses.

The new software framework encompassing these SpTTN kernels, SpTTN-Cyclops, is an extension of the 
CTF~\cite{ctf} library for sparse/dense distributed tensor contractions.
CTF provides routines for mapping sparse or dense tensor data to multidimensional processor grids and redistributing data between any pair of grids.
Given a mathematical description of a tensor and a sets of contractions, CTF automatically finds a contraction path (sequence of pairs of tensors to contract) and performs each contraction in parallel on a suitable grid.
SpTTN-Cyclops instead simultaneously contracts the sparse tensor with all dense tensors in the tensor network, forgoing construction of large (sparse) intermediate tensors required by the CTF method.
This all-at-once contraction method has been shown to be efficient in theory and practice for some specific SpTTN kernels such as MTTKRP~\cite{ballard_mttkrp, ballard_mem_ind_mttkrp, SPLATT, CSF}.

The all-at-once contraction approach allows SpTTN-Cyclops to keep the sparse tensor data in place, and rely on existing CTF routines for communication of the other operands.
Locally, each processor must then simply execute a loop nest for a smaller SpTTN of the same type.
Our framework leverages the new SpTTN loop nest enumeration and search algorithms to select the best choice of loop nest, which is not possible with any previously existing library.
To achieve good performance for the innermost loops, we leverage the Basic Linear Algebra Subroutines (BLAS)~\cite{BLAS}, whenever possible, and incorporate this into our cost function.
A similar technique has been used in Mosaic~\cite{mosaic}, a sparse tensor algebra compiler that demonstrates the benefits of binding tensor sub-expressions to external functions of other tensor algebra libraries and compilers.

We evaluate our framework against the single node performance of TACO and SparseLNR, and the distributed memory implementation of CTF. We also compare SpTTN-Cyclops with the state-of-the-art specialized implementation of one of the SpTTN kernels (SPLATT \cite{SPLATT}).
Our results demonstrate that we achieve higher performance or close to SPLATT's specifically tuned implementation of one of the kernels. 
We outperform all three generalized frameworks (TACO, SparseLNR, and CTF) by orders of magnitude. Across some of the kernels, we achieve speedups in the range of 2 to 100x when compared to these generalized frameworks.
We show strong scaling results in the distributed memory setting using tensors of various dimensions and sparsity. We also enable the computation of some of these kernels on larger tensor inputs for which the other frameworks run out of memory.

\section{Background}
\begin{figure*}
  \centering
  \begin{subfigure}[b]{0.23\textwidth}
      \centering
      \includegraphics[width=\textwidth]{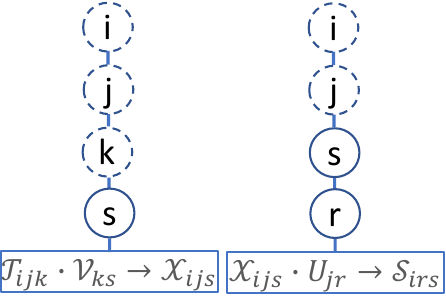}
      \caption{Each pairwise contraction has an independent loop nest.}
      \label{fig:ig_independent}
  \end{subfigure}
  \hfill
  \begin{subfigure}[b]{0.23\textwidth}
      \centering
      \includegraphics[width=\textwidth]{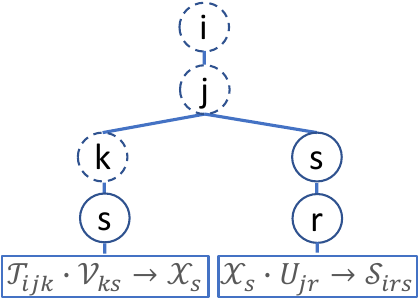}
      \caption{Vertices $i$ and $j$ are fused across the two pairwise contractions. \conditionalPara{The order of the intermediate tensor $\tsr{X}$ is reduced by 2 since $i$ and $j$ are no longer buffered.}}
      \label{fig:ig_fused}
  \end{subfigure}
  \hfill
  \begin{subfigure}[b]{0.23\textwidth}
      \centering
      \includegraphics[width=\textwidth]{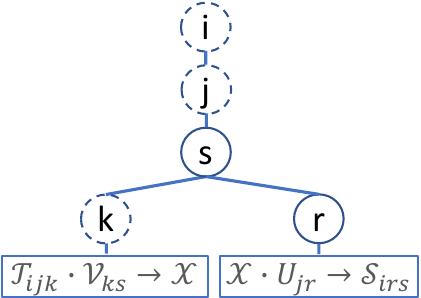}
      \caption{Vertices $i$, $j$ and $s$ are fused across the two pairwise contractions. \conditionalPara{The intermediate tensor $X$ is a scalar as none of the indices are buffered.}}
      \label{fig:ig_loop_order}
  \end{subfigure}
  \hfill
  \begin{subfigure}[b]{0.23\textwidth}
      \centering
      \includegraphics[width=\textwidth]{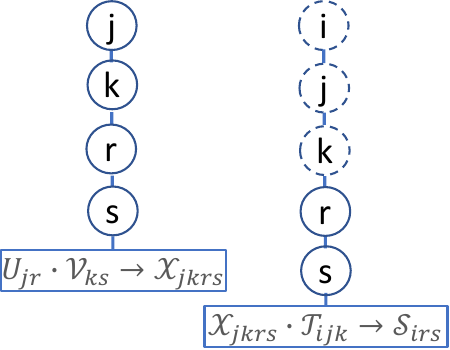}
      \caption{None of the vertices can be fused.}
      \label{fig:ig_pairwise_seq}
  \end{subfigure}
     \caption{Graphs illustrating loop nests for computing an order 3 TTMc kernel. Sparse loops are shown as dotted vertices.}
     \label{fig:ig}
\end{figure*}

\begin{table*}
\noindent\begin{minipage}{.28\textwidth}
\begin{lstlisting}[aboveskip=0em, belowskip=-1em, label={lst:ttmc_pairwise}, caption={TTMc kernel computed via pairwise contractions.}]
T_csf = CSF(${\tsr{T}\_ijk}$)
$\tsr{X}$ = 0
for (i,T_i) in T_csf:
  for (j,T_ij) in T_i:
    for (k,t_ijk) in T_ij:
      for s in range(S):
        $\tsr{X}$[i,j,s] += t_ijk * $\tsr{V}$[k,s]
for (i,T_i) in T_csf:
  for (j,T_ij) in T_i:
    for s in range(S):
      for r in range(R):
        $\tsr{S}$[i,r,s] += $\tsr{X}$[i,j,s] * $\tsr{U}$[j,r]
\end{lstlisting}
\end{minipage}
\hfill 
\noindent\begin{minipage}{.35\textwidth}
\begin{lstlisting}[aboveskip=0em, belowskip=0em, label={lst:ttmc_fusion},caption={TTMc kernel computed using the factorize-and-fuse approach. A single loop nest of $i$,$j$ is used to iterate over both the pairwise contractions.}]
T_csf = CSF(${\tsr{T}\_ijk}$)
for (i,T_i) in T_csf:
  for (j,T_ij) in T_i:
    $\tsr{X}$ = 0 // reset intermediate tensor
    for (k,T_ijk) in T_ij:
      for s in range(S):
        $\tsr{X}$[s] += t_ijk * $\tsr{V}$[k,s]
    for s in range(S):
      for r in range(R):
        $\tsr{S}$[i,r,s] += $\tsr{X}$[s] * $\tsr{U}$[j,r]
\end{lstlisting}
\end{minipage}
\hfill 
\noindent\begin{minipage}{.32\textwidth}
\begin{lstlisting}[aboveskip=0em, belowskip=0em, label={lst:ttmc_loop_order},caption={TTMc kernel computed using the factorize-and-fuse approach. Indices $i$,$j$,$s$ are fused.}]
T_csf = CSF(${\tsr{T}\_ijk}$)
for (i,T_i) in T_csf:
  for (j,T_ij) in T_i:
    for s in range(S):
      $\tsr{X}$ = 0 // reset intermediate tensor
      for (k,T_ijk) in T_ij:
        $\tsr{X}$ += t_ijk * $\tsr{V}$[k,s]
      for r in range(R):
        $\tsr{S}$[i,r,s] += $\tsr{X}$ * $\tsr{U}$[j,r]
\end{lstlisting}
\end{minipage}
\end{table*}

\subsection{Tensor Notation}
We use calligraphic letters to denote tensors, e.g., $\tsr{T}$. 
An order $N$ tensor corresponds to an $N$-dimensional array. 
We denote elements of tensors in parenthesis, e.g., $\tsr{T}(i,j,k,l)$ for an order 4 tensor $\tsr{T}$. 
The indices that do not appear in the output tensor are considered to be summed (contracted). 
\conditionalPara{For example, matrix-matrix multiply that is contracted over index $k$ is denoted as 
$\tsr{A}(i,j) = \sum_{k} \tsr{B}(i,k) \cdot \tsr{C}(k,j)$.}
We use capitalized letters to denote the dimensions of the respective indices. For example, the dimension of index $i$ in $\tsr{A}(i,j)$ is denoted as $I$.

\subsection{Tensor Sparsity and Sparse Storage}
\label{sec:csf_tree}
We use one of the most common ways to store sparse tensors, the Compressed Sparse Fiber (CSF) format \cite{CSF}.
We refer to the total number of nonzero elements of a tensor $\mathcal{T}$ as $\nnz(\mathcal{T})$.
For a sparse tensor $\mathcal{T}$ with $d$ dimensions of size $I_1,\ldots,I_d$, we represent the number of non-zeroes in the $k$th level of the CSF tree for $\mathcal T$ (with the first index being at the root) as $\nnz^{(I_{1}\cdots I_{k})}(\mathcal T)$.
Equivalently, this nonzero count may be obtained by considering the number of nonzeros in a reduced tensor obtained by summing away the remaining modes, i.e., $\nnz^{(I_{1} \cdots I_{k})}(\mathcal T) = \nnz(\mathcal S)$, where
$\mathcal S(i_{1},\ldots,i_k) = \sum_{i_{k+1},\ldots,i_d} |\mathcal{T}(i_1,\ldots, i_d)|$.
\subsection{Tensor Decomposition and Completion Algorithms}
\label{sec:tensor_decomp_completion}

Tensor decomposition \cite{kolda_tensor_review} and completion \cite{smith_tensor_completion} refer to the problem of decomposing a tensor into a combination of smaller tensors and estimating missing or incomplete values in a tensor, respectively. 
\conditionalPara{These techniques are often used for feature extraction, compression, or when the tensor data is partially observed or noisy.
Tensor decomposition and completion 
can be seen as generalization of matrix decomposition and completion, respectively.}

The algorithms for both tensor decomposition and completion
focus on a single sparse tensor (the input dataset) and require computations that factorize or update the tensor by contracting it with several smaller dense tensors (representing the decomposition). 
These computations, which we refer to as \textit{kernels},
account for a significant percentage of the overall execution of an algorithm.
They have been the focus of high-performance implementations and are typically available as specialized libraries \cite{SPLATT, HyperTensor, AdaTM, MTTKRP_Emerging_Arch, MTTKRP_Blocking, parti}.
We list some of the kernels below and describe their existing implementations in the Section \ref{sec:kernel_implementation}. 
  \\ 1. A standard approach to compute the Canonical Polyadic (CP) decomposition \cite{cp} of a tensor is the alternating least squares (ALS) algorithm. Matricized-Tensor times Khatri-Rao Product (MTTKRP) is a key kernel in computing CP-ALS and is the main bottleneck \cite{DFacTo,GigaTensor,CSTF},
  \begin{align} 
  \tsr{A}(i,a) = \sum_{j,k} \tsr{T}(i,j,k) \cdot \tsr{B}(j,a) \cdot \tsr{C}(k,a).
  \label{eqn:mttkrp}
  \end{align}
   2. For Tucker decomposition \cite{tucker}, the analogous to ALS is the higher-order orthogonal iteration (HOOI) algorithm. The primary kernel in HOOI is the tensor-times-matrix chain (TTMc) \cite{accelerate-tucker,ptucker},
   \begin{align} 
   \tsr{S}(i,r,s) = \sum_{j,k} \tsr{T}(i,j,k) \cdot \tsr{U}(j,r) \cdot \tsr{V}(k,s).
   \label{eqn:ttmc}
   \end{align}
   3. A common generic multi-tensor kernel in tensor completion is the tensor-times-tensor-product (TTTP) \cite{SINGH2022269}. TTTP generalizes the sampled dense-dense matrix multiplication (SDDMM) kernel \cite{sddmm-tttp_1,sddmm-tttp_2}, and is also useful for CP decomposition of sparse tensors,
  \begin{align} 
  \tsr{S}(i,j,k) = \sum_{r} \tsr{T}(i,j,k) \cdot \tsr{U}(i,r) \cdot \tsr{V}(j,r) \cdot \tsr{W}(k,r).
  \label{eqn:tttp}
  \end{align}
  Note that $\tsr{S}$ has the same sparsity pattern as that of $\tsr{T}$.
  \\ 4. Tensor-Times-Tensor-chain (TTTc) kernel used in sparse tensor train decomposition \cite{stto} to decompose a higher order sparse tensor using first-order optimization methods,
  \begin{align}
  \tsr{Z}(e,n) = \sum_{\mathclap{i,j,k,l,m,n,a,b,c,d}} \tsr{T}(i,j,k,l,m,n) \cdot \tsr{A}(i,a) \cdot \tsr{B}(a,j,b) \notag \\
  \cdot \tsr{C}(b,k,c) \cdot \tsr{D}(c,l,d) \cdot \tsr{E}(d,m,e).
  \end{align}
We restrict attention to sparse tensor kernels where the output is dense or has the exact same sparsity as the sparse input tensor. This precludes some common kernels, such as tensor times matrix (TTM) \cite{spttm} and contraction of two sparse tensors (e.g., SpGEMM \cite{spgemm}), since these generally produce a sparse output.
\vspace{-0.05in}
\subsection{Computation of Tensor Kernels in Decomposition and Completion Algorithms}
\label{sec:kernel_implementation}
In this section we describe the existing approaches to compute the kernels listed in Section \ref{sec:tensor_decomp_completion}.

\subsubsection{Unfactorized Contraction}
\ \\
To compute a kernel, we can iterate over all the indices and simultaneously contract all the input tensors in the innermost loop.
We refer to this approach as \textit{unfactorized}.
This unfactorized loop nest has a depth equal to the number of unique indices.
For example, consider an order 3 TTMc kernel in Equation \ref{eqn:ttmc}. 
The number of operations is $3 \nnz(\mathcal{T})\cdot R\cdot S$ to leading order.
In compiler driven frameworks such as TACO \cite{TACO} and COMET \cite{comet}, the schedule generated by default is unfactorized.

The unfactorized approach is optimal in cost for computing certain kernels.
For example, the MTTKRP kernel in Equation~\ref{eqn:mttkrp} can be computed using the unfactorized approach with an optimal loop depth of 4.
But this approach is asymptotically suboptimal for many other kernels, such as the TTMc.

\subsubsection{Pairwise Contraction}
\label{sssection:pairwise_contraction}
\ \\
A kernel can be computed with minimal asymptotic complexity (loop depth) by contracting the input tensors pairwise. We refer to this approach as \textit{pairwise contraction}. 
It is typically used in libraries designed for dense tensor contractions, such as CTF~\cite{ctf}, Tensor Computation Engine (TCE)~\cite{TCE}, and DEinsum~\cite{Deinsum}.
For example, consider the TTMc kernel in Equation \ref{eqn:ttmc}. One way in which the tensors can be contracted pairwise is to first contract $\tsr{T}$ with $\tsr{V}$, and then its result with $\tsr{U}$.
Each 
pairwise contraction has an independent loop nest as shown in Listing \ref{lst:ttmc_pairwise}.
Both the loop nests have a depth of 4, and the computational cost is $2\nnz(\mathcal{T})\cdot S + 2\nnz^{(IJ)}\cdot S \cdot R$ to leading order.
Even though an unfactorized approach for computing the MTTKRP kernel (Equation \ref{eqn:mttkrp}) has an optimal loop depth, up to a third of the operations can be eliminated by using pairwise contraction.
The unfactorized approach requires $3\cdot \nnz(\mathcal{T})\cdot A$ scalar operations, while the pairwise approach requires $2\nnz^{(IJK)}(\mathcal T)\cdot A + 2\nnz^{(IJ)}\cdot A$ operations.
\conditionalPara{This cost is achieved by contracting the tensors $\tsr{T}$ and $\tsr{C}$ pairwise, and then contracting the result with $\tsr{B}$.}
\if showStriked
\begin{align}
  \tsr{X}(i,j,a) = \sum_{k} \tsr{T}(i,j,k) \cdot \tsr{C}(k,a), \notag \\ 
  \tsr{A}(i,a) = \sum_{j} \tsr{X}(i,j,a) \cdot \tsr{B}(j,a),
\label{eqn:pairwise_mttkrp}
\end{align}
\fi
\conditionalPara{,where $\tsr{X}$ is an intermediate tensor}

For contractions involving only dense tensors, the pairwise approach can provide an optimal schedule.
But for sparse tensors, whose dimensions are often large, this approach can lead to unmanageable memory requirements for storing dense intermediate tensors.
In practice, pairwise contraction with sparse storage of such an intermediates has been observed to be much slower than hand-tuned or even unfactorized implementations for SpTTN kernels~\cite{SINGH2022269}.

\subsubsection{Factorize-and-Fuse}
\label{sec:loop_fusion}
\hfill\\
The size of the intermediate tensors can be reduced by loop fusion. 
Loop nests that share a common index can be nested together with an outer loop that iterates over the shared index.
The loop nests that compute the pairwise contractions in 
Listing \ref{lst:ttmc_pairwise}
can be fused together as shown in Listing \ref{lst:ttmc_fusion}.
We refer to this approach as \textit{factorize-and-fuse}.
A single loop nest of $i$,$j$ is used to iterate over both the pairwise contractions and hence the indices are not buffered.
The computation cost remains the same as in the pairwise case (in fact, the same set of operations is computed).
The size of the intermediate tensor $\tsr{X}$ is reduced from $I \times J \times S$ to $S$. 
Specialized libraries for some of these kernels use a similar approach in their hand-tuned implementations \cite{SPLATT, HyperTensor, AdaTM, MTTKRP_Emerging_Arch, MTTKRP_Blocking, parti, GigaTensor,accelerate-tucker}.

\externaldocument{background}
\section{SpTTN Kernels}
\label{sec:spttn_kernels}
In Section \ref{sec:tensor_decomp_completion}, we listed several kernels for tensor decomposition and completion.
We now aim to define these generally.
To motivate this definition, consider any tensor decomposition or completion of tensor $\tsr{T}$ given by a model $\tilde{\tsr{T}}$ composed of dense tensors $\tsr{A}_1,\dots, \tsr{A}_n$ (factors), the objective function, denoted by $f$ is expressed as,
\begin{align*}
f(\tsr{A}_1,\dots,\tsr{A}_n) = \|\tsr{T} - \tilde{\tsr{T}}(\tsr{A}_1,\dots,\tsr{A}_n)\|_F^2.
\end{align*}
The optimization methods generally leverage all or parts of the gradient of the residual error norm, which yields a contraction of the sparse tensor, with subsets (all but one of the) tensors from the decomposition. 
\if showStriked
i.e. the gradient of $f$ will be
\begin{align*}
\nabla f(\tsr{A}_1,\dots,\tsr{A}_n) = [-2\tsr{T}\cdot \tilde{\tsr{T}}(\varnothing,\tsr{A}_2,\dots,\tsr{A}_n) + \\ 
  \tilde{\tsr{T}}(\varnothing,\tsr{A}_2,\dots,\tsr{A}_n) \cdot \tilde{\tsr{T}}(\varnothing,\tsr{A}_2,\dots,\tsr{A}_n); \\
        -2\tsr{T} \cdot \tilde{\tsr{T}}(\tsr{A}_1,\varnothing,\tsr{A}_3,\dots,\tsr{A}_n) + \\
        \tilde{\tsr{T}}(\tsr{A}_1,\varnothing,\tsr{A}_3,\dots,\tsr{A}_n) \cdot \tilde{\tsr{T}}(\tsr{A}_1,\varnothing,\tsr{A}_3,\dots,\tsr{A}_n); \cdots ].
\end{align*}
The terms involving $\tsr{T}$ are cost-dominant.
\fi
The terms involving $\tsr{T}$ when computing the gradient of $f$ are cost-dominant.
Similarly, when computing the residual error ($\rho$) for tensor completion, which is often employed, e.g., in coordinate descent methods, the terms involving the sparse tensor are cost-dominant.
$\rho = \tsr T - \Omega \ast \tilde{\tsr T}(A_1,\ldots,A_n)$,
where $\ast$ is the Hadamard (pointwise product), the sparse tensor $\Omega$ represents the set of observed entries in the input tensor $\tsr{T}$ and $\tilde{\tsr{S}}$ is the output tensor obtained by contracting $\Omega$ with a network of dense factors.

In general, we define an \textit{SpTTN kernel} as a contraction of a sparse tensor with a set of dense tensors resulting in an output with a dense representation or a sparse tensor with the same sparsity as the sparse input tensor.
Hence, in any SpTTN, a subset of the indices in the contraction has a fixed/known sparsity pattern (associated with the input sparse tensor), while the remaining indices iterate only over dense tensors.
We generally assume the dense tensors in the SpTTN are fairly small (in comparison to the~input~sparse~tensor).

\externaldocument{background}
\externaldocument{new_cp_lo}
\subsection{Loop Nests and Loop Nest Forests}
\label{sec:loop_nests}

The computation of a tensor contraction generally involves loop nests of some form. 
Any loop nest can be represented by an ordered tree or forest, each vertex of which is a loop, and its ordered children are the loop nests contained directly in that loop.
Each leaf corresponds to a contraction term (a pair of tensors contracted together). For example, the loop nest in Listing \ref{lst:ttmc_pairwise} is represented by the tree in Figure \ref{fig:ig_independent}. We refer to a tree with a single leaf as a \textit{path graph}.
A similar representation is used in TACO \cite{TACO}.

\begin{figure}[h]
\centering
\includegraphics[width=.5\textwidth]{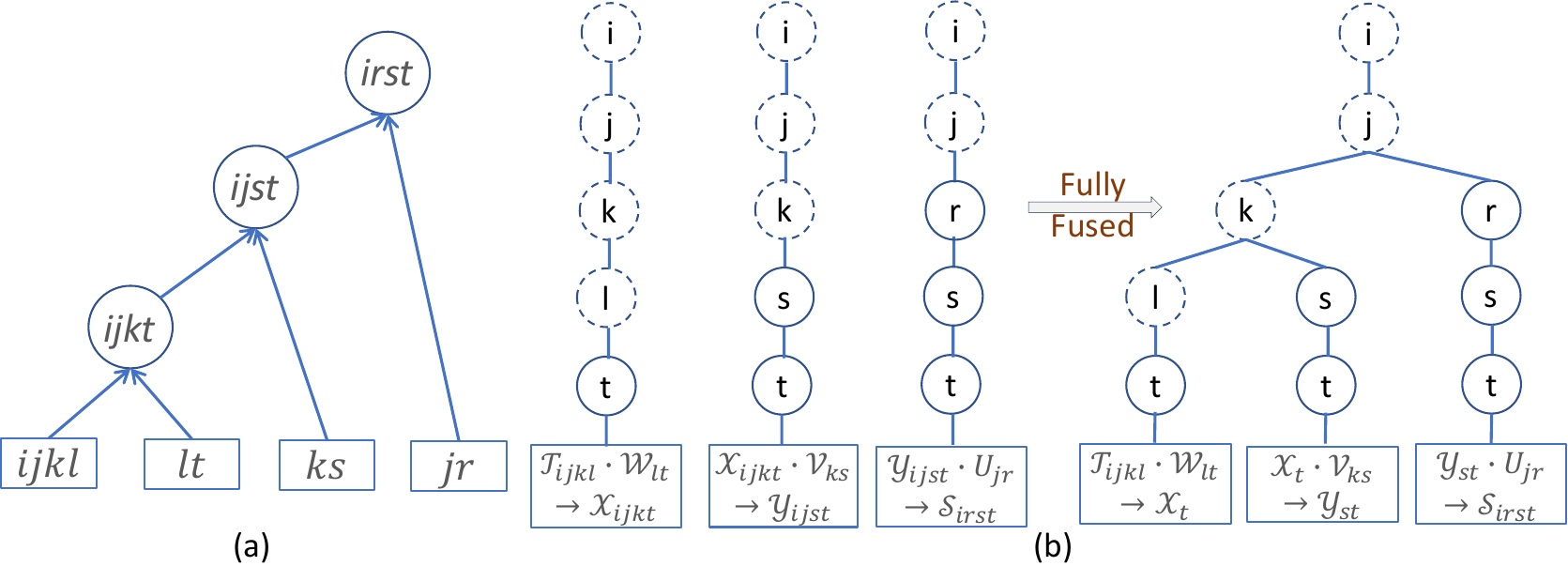}
\caption{
An order 4 TTMc kernel $\tsr{S}_{irst} = \tsr{T}_{ijkl}\cdot \tsr{U}_{jr}\cdot \tsr{V}_{ks}\cdot \tsr{W}_{lt}$, where (a) represents the contraction path tree ($T$) with $L = ((ijkl,lt,ijkt), (ijkt,ks,ijst), (ijst,jr,irst))$, and (b) shows the path graphs corresponding to the contraction path terms, fused to obtain a fully fused loop nest tree.
}
\label{fig:fully_fused}
\vspace{-0.1in}
\end{figure}

The leaves of the loop nest tree define the order in which the contraction terms are executed. We refer to this order as the \textit{contraction path}. A contraction path for a kernel is valid if we can obtain the output tensor by executing the contraction terms in the order specified by it.
\begin{definition}[Contraction Path]
For a contraction of $N+1$ tensors, a contraction path is given by a depth-first postordering of a $2N+1$-node binary tree $T$ where the $N+1$ leaves are the input tensors, and each internal node corresponds to the contraction of a pair of input tensors and/or intermediates, so all non-leaf nodes have exactly two children.
We represent a contraction path by the tree $T$ and an ordered collection of index set 3-tuples, $L = (L_1,\dots,L_N)$, where each $L_i$ contains the indices of the tensor operands and output at each of the $N$ internal~tree~nodes.
\end{definition}
Note that while a contraction path is defined above based on a tree, this tree is different tree from 
a loop nest tree.
In a loop nest tree, each node corresponds to a loop and each leaf is a term in the contraction path.
Hence, a node in the contraction path tree corresponds to a leaf in the loop nest tree.
Figure \ref{fig:fully_fused}(a) shows a contraction path tree for an order 4 TTMc kernel.

In a valid loop nest forest, all indices in a contraction term should be loop indices on the path 
between the corresponding leaf and the tree root, and the path should contain no additional or repeated indices.
We refer to this order of loop indices as the \emph{loop order} of the contraction term.
For example, in Figure \ref{fig:fully_fused}(b), the loop order of the first term, 
$\tsr{T}_{ijkl}\cdot \tsr{W}_{lt}\rightarrow \tsr{X}_{ijkt}$, is $(i,j,k,l,t)$.
If a vertex has multiple leaves in its subtree, the loop associated with that vertex contains all the contraction terms in that subtree.
\begin{definition}[Loop Order]
A loop order for a contraction path $(T,L)$, $L = (L_1,\ldots,L_N)$ is defined by an ordered collection $A=(A_1,\dots,A_N)$, where each $A_i$ is an ordered collection of the union of the indices in the 3 index sets contained in $L_i$.
\end{definition}
\vspace{-5pt}
We say a loop nest tree is fully-fused if no vertex has two consecutive children that correspond to the same index.
A fully-fused loop nest and the corresponding tree 
is obtained by fusion of the path graphs (loop nests) corresponding to each term in $A$.
In Figure \ref{fig:fully_fused}(b), the path graphs corresponding to the contraction path terms are fused to obtain a fully-fused loop nest tree.
A loop nest forest is fully-fused if adding a dummy vertex and connecting it to all roots in the forest yields a fully-fused loop nest tree.


\conditionalPara{
Figure \ref{fig:ig} shows loop nest forests for TTMc with an order 3 sparse tensor (Equation \ref{eqn:ttmc}). 
A pairwise contraction of the kernel has two path graphs (independent trees, one for each contraction term), and the contraction path gives the order in which these terms are computed. This loop nest forest is equivalent to the loop nest presented in Listing \ref{lst:ttmc_pairwise}.
Similarly, loop nest for an unfactorized contraction of the kernel can be represented by a path graph. The leaf corresponds to contracting the input tensors unfactorized to obtain the output tensor.
The factorize-and-fuse approach in Listing \ref{lst:ttmc_fusion} yields a fully-fused tree as shown in Fig \ref{fig:ig_fused}.}

\subsection{Intermediate Tensors}
\label{subsec:intm_size}
Every contraction term except the last, writes its output to an intermediate tensor (buffer).
Let the term that generates an intermediate tensor and the term that consumes it be $L_x$ and $L_y$, respectively.
The indices of the intermediate tensor $\tsr{I}_{L_xL_y}$ are given by
\vspace{-0.035in}
\begin{align}
\label{eqn:intermediate_indices}
  \texttt{inds}(\tsr{I}_{L_xL_y}) = (\texttt{inds}(L_x) \cap \texttt{inds}(L_y)) \setminus S,
\end{align}
where $S$ is the set of common ancestors of the two terms in the loop nest graph.


\subsection{Contraction Path and Loop Order}
\label{sec:cp_io}
The contraction path affects the asymptotic complexity (loop depth) and memory requirements (intermediate tensor sizes) of the computation.
For example, consider the various ways to compute the TTMc kernel as shown in Figure \ref{fig:ig}. In one of the chosen contractions paths, tensors $\tsr{T}$ and $\tsr{V}$ are contracted first and the result is then contracted with $\tsr{U}$. The computation has a maximum loop depth of 4 (Figures~\ref{fig:ig_independent}, \ref{fig:ig_fused} and \ref{fig:ig_loop_order}). 
A different contraction path of the same kernel, where tensors $\tsr{U}$ and $\tsr{V}$ are contracted first and then the result is contracted with $\tsr{T}$, yields a maximum loop depth of 5~(Figure~\ref{fig:ig_pairwise_seq}).
%

Similarly, for a given contraction path, the ordering of vertices in the path graphs before fusing them, affects the intermediate tensor sizes and other cost metrics of interest. 
In the previous example of the TTMc kernel, consider the iteration graph in Figure \ref{fig:ig_independent} and its fully-fused variant in Figure \ref{fig:ig_fused}.
Indices $i$, $j$ and $s$ are common across the two trees in the iteration graph.
We are able to fuse vertices $i$ and $j$ but not $s$ (loop order in the first path graph is $(i,j,k,s)$). This results in an intermediate tensor of size $S$ in Figure~\ref{fig:ig_fused} (see Listing~\ref{lst:ttmc_fusion}).
But if the loop order in the first path graph is $(i,j,s,k)$, we can fuse vertices $i$, $j$ and $s$ in the iteration graph and obtain a fully-fused loop nest tree with an intermediate tensor of size 1 (scalar) (see Figure \ref{fig:ig_loop_order} and its corresponding loop nest in Listing \ref{lst:ttmc_loop_order}).
In the next section, we seek to find cost-optimal loop nests for a given SpTTN kernel, where the cost is defined by a cost model, for example, the intermediate tensor size.

\if showStriked
\begin{lstlisting}[float=h,aboveskip=0em, belowskip=0em, label={lst:ttmc_contraction_order},caption={TTMc kernel computed using the factorize-and-fuse approach. Tensors $\tsr{U}$, $\tsr{V}$ are contracted first to produce an intermediate tensor $\tsr{X}$. The main tensor $\tsr{T}$ is then contracted with $\tsr{X}$ to compute the output tensor $\tsr{S}$.}]
T_csf = CSF(${\tsr{T}\_ijk}$)
// $j$ and $k$ in the first loop nest are treated as dense loops
$\tsr{X}$ = 0
for j in range(J):
  for k in range(K):
    for r in range(R):
      for s in range(S):
        $\tsr{X}$[j,k,r,s] += $\tsr{U}$[j,r] * $\tsr{V}$[k,s]
for (i,T_ijk) in T_csf:
  for (j,T_ij) in T_i:
    for (k,T_ijk) in T_ij:
      for r in range(R):
        for s in range(S):
          $\tsr{S}$[i,r,s] += $\tsr{X}$[j,k,r,s] * t_ijk
\end{lstlisting}
\fi

\section{Finding Optimal SpTTN Kernels}
\label{sec:algorithm}
To determine an efficient loop nest for an SpTTN kernel, we first present an approach to enumerate fully-fused trees and later present efficient algorithms to find an optimal tree for simple cost metrics. 


\subsection{Enumeration of Loop Nests}
We seek to find cost-optimal loop nests for a given SpTTN kernel by enumerating only fully-fused loop nest forests and restrict our attention to dense multidimensional buffers (intermediate tensors).
We decouple the enumeration into two steps: (1) enumeration of valid contraction paths for a given set of tensors and (2) enumeration of loop orders in the path graphs for a given contraction path.

\subsubsection{Enumeration of Contraction Paths}
\hfill\\
Let the number of input tensors in the SpTTN be $n$.
To enumerate contraction paths, we employ a function to pick and contract all combinations of two tensors from the list of input tensors. We then recurse over a new list constructed by replacing each pair of contracted tensors with the contraction output. This approach has been studied in the context of finding an optimal contraction path for dense tensor networks \cite{faster_optimal_contraction}.
The cost can be analyzed by the recurrence relation, $T(n) = {n \choose 2} \cdot T(n-1)$ and $T(2) = 1$ (when there are two tensors to contract). The number of valid contraction paths for $n$ tensors is $O \left( \frac{(n!)^2}{n \cdot 2^n} \right)$.

In~\cite{ilya-cameron}, dynamic programming is used to find the cost-optimal contraction path (tree) given a fixed order of dense tensors to be contracted. 
This approach is analogous and complementary to our work of finding a cost-optimal loop nest tree for a given contraction path, which we present in Section~\ref{sec:cost_optimal_loop_nests}.

\subsubsection{Enumeration of Loop Orders for a Given Contraction Path}
\hfill\\
For a given contraction path, we construct a path graph for each term by picking a loop order for that term.
The path graphs are then fused to obtain a fully fused loop nest tree. Each choice of loop order yields a different fused loop nest.

Let the set of indices in the $i$th term be $I_i$.
The set of indices in the SpTTN is given by $I = \bigcup_{i=1}^{n-1} I_i$.
We do an exhaustive search by enumerating all loop orders independently for each path graph and then considering all possible combinations of these orders.
Since we do not allow any repeated indices in our path graphs, the loop nests generated in such an enumeration are unique and span all the possible loop nests for a given contraction path.
The cardinality of this exhaustive search is given by $\prod_{i=1}^{n-1} |I_i|!$.
We later restrict the search to only those loop orders that are consistent with the order of the indices of the sparse tensor, so if a term involves $k$ sparse indices, the number of possible orders for the term is only $|I_i|!/k!$.

\subsubsection{Upper Bound on Loop Nests}
\hfill\\
For a given SpTTN, the number of loop nests we enumerate has an upper bound given by the product of the number of contraction paths and the number of loop orders for a given contraction path, i.e.,
$O\Big(\frac{(n!)^2\cdot \prod_{i=1}^{n-1} |I_i|!}{n\cdot 2^n}\Big)$.
In the following section, we present a dynamic programming algorithm to prune the search space for loop order enumeration.

\subsection{Algorithm to Find Cost-Optimal Loop Nests}
\label{sec:cost_optimal_loop_nests}
Enumeration enables autotuning, but for analytic metrics of performance such as buffer size, more efficient search schemes are possible.
Different contraction paths yield different fully-fused loop nests, hence we focus our attention to enumeration and search of loop nests for a particular contraction path.
In TCE~\cite{tce_dp,tce_dp_tiling}, dynamic programming is used to find the cost-optimal loop nest for dense tensor contractions, with one of the cost metrics being the intermediate tensor size.
Our efficient search algorithm also employs dynamic programming, after decoupling order of terms from the tree structure.
Given a fixed contraction path order (or a subsequence of the terms, which defines a subproblem), we seek to find a loop nest tree that minimizes 
a chosen cost metric.



\externaldocument{background}

We introduce a peeling primitive for fully fused loop nests to formally define the tree structure.
Peeling a fully fused loop nest removes the first outermost loop nest.
In a fully-fused loop nest, the outermost loop should iterate over an index that appears in the first contraction, and include within it all subsequent contractions in the contraction path order until one does not include the index.
\begin{definition}[Peeling of Loop Order]
Given loop order $A=(A_1,\ldots, A_N)$, choose $r\in\{1,\ldots, N\}$ to be the largest such that $A_1[1]=A_2[1]=\cdots =A_r[1]$.
Peeling $A$ yields two loop orders $A^{(1)}=(A_1[2:\ ],\ldots, A_r[2:\ ])$ and $A^{(2)}=(A_{r+1},\ldots, A_{N})$ (where $A_x[2:\ ]$ denotes the subsequence of all elements in $A_x$ except the first element and is omitted if $A_x$ has size 1).
\end{definition}
The loop nest tree or forest can then be constructed from the representation $A=(A_1,\ldots, A_N)$ by peeling $A$ iteratively and adding vertices for the two resulting loop orders (if not empty).
\begin{definition}[Fully-fused Loop Nest Forest]
Given an loop order $A=(A_1,\ldots, A_n)$, the corresponding fully-fused loop nest forest $\mathcal{F}(A)=(V,E)$ is constructed as follows.
Initialize $V$ as one vertex corresponding to loop index $A_1[1]$, then apply peeling iteratively.
At each peeling step, add vertices to $V$ for $A^{(1)}$ and $A^{(2)}$ (unless they are zero-sized) connecting $A^{(1)}$ to the vertex representing $A$ and $A^{(2)}$ to its parent (if any).
\end{definition}
To work with analyzing loop nest forests, it also helps to think about the effect of peeling the loop order on the loop nest tree associated with the loop order.
\begin{definition}[Peeling of Fully-fused Loop Nest Tree]
Given a loop nest loop order $A$ for contraction path $(T,L)$ and the corresponding fully-fused loop nest tree $\mathcal{F}(A)=(V,E)$, peeling removes the root vertex (index $r$) of the tree.
If the root has $k$ children, the resulting independent subtrees are associated with loop orders $B^{(1)},\ldots, B^{(k)}$, each of which computes a contraction path for distinct subsets of terms $L^{(1)},\ldots, L^{(k)}\subseteq \hat{L}$, where $\hat{L}$ is defined by removing the index $r$ from all index sets in $L$.
The contraction path tree for the $i$th loop order, $T^{(i)}$, is given by removing all vertices from $T$ except those corresponding to terms computed in $L^{(i)}$ and their children (inputs).
\end{definition}

\subsubsection{General Cost Function}
\label{ssec:cost_function}
\hfill\\
In general, the execution time of a particular fully-fused loop nest tree may depend on architecture or data sparsity in ways that are impractical to fully model and require enumeration and execution.
On the other hand, for a simple cost function, e.g., computational cost\footnote{Since the same contraction path is being considered, all fully-fused loop nest trees have the same asymptotic complexity in tensor size, but order and fusion have an affect on lower-order cost terms.} or intermediate buffer size, the search space can be explored more systematically and efficiently.
However, more sophisticated cost functions, which take into account metrics such as cache-efficiency or parallelizability are also of clear interest.
We now define a class of functions which we can optimize efficiently, requiring separability of cost according to the structure of the loop nest tree.
\externaldocument{algorithm}
\begin{definition}[Tree-separable Cost Function]
\label{def:treesep}
Consider a loop nest order $A$ for a contraction path $(T,L)$.
Let $B^{(1)},\ldots, B^{(k)}$ be the loop nest orders for subtrees obtained after peeling root $r$ of tree $\mathcal{F}(A)$ and $(T^{(i)},L^{(i)})$ be the corresponding contraction path for each $B^{(i)}$.
A cost function $f_{\phi,\oplus}$ for this loop nest is tree-separable if it satisfies,
\begin{align*}
f_{\phi,\oplus}(T,L,A) = \phi_{T,L,r}\Big( &f_{\phi,\oplus}(T^{(1)},L^{(1)},B^{(1)})\oplus \cdots \oplus \\
& f_{\phi,\oplus}(T^{(k)},L^{(k)},B^{(k)})\Big),
\end{align*}
where $\phi_{T,L,r}:R_+\to R_+$ is nondecreasing and $\oplus$ is an associative semigroup operator on $R_+$ that is nondecreasing in both variables.
If $\mathcal{F}(A)$ is a forest, $f_{\phi,\oplus}(T,L,A)$ is given by combining the costs of the independent trees with $\oplus$.
\end{definition}
This definition is quite general as $\phi$ is parameterized by the contraction path, and so could be defined at each loop level with full information of the indices/terms involved in the nested loops it contains.
At the same time, we observe that $f$ can be evaluated on $A$ recursively, as $\phi$ does not depend on all of $A$, but only the contraction path and the root vertex of $\mathcal{F}(A)$.
We could also allow the same parameterization for $\oplus$ without overhead in search complexity, but do not do so for simplicity and due to lack of need.
\subsubsection{Maximum Buffer Size}
\label{ssec:cost_function_mbs}
\hfill\\
We now provide a tree-separable cost function to compute the maximum dimension of the intermediate tensors/buffers produced in the execution of a fully fused loop nest. We interchangeably use the terms intermediate tensor and buffer.
\begin{definition}[Cost Function for Maximum Buffer Dimension]
Consider a fully fused loop nest tree $\mathcal{F}(A)$ for loop order $A$ with contraction path $(T,L)$, where  $T=(V,E)$.
Let $B^{(1)},\ldots, B^{(k)}$ be the loop nest orders for subtrees obtained after peeling $\mathcal{F}(A)$ and $(T^{(i)},L^{(i)})$ be the corresponding contraction path for each $B^{(i)}$.
Let $Z\subseteq E$ be the set of edges in the contraction path (oriented towards the root) connecting a node that corresponds to a term $L_u\in B^{(i)}$ to another,  $L_v\in B^{(j)}$ with $i\neq j$. 
The maximum buffer dimension used in the fully fused loop nest is given by $f_{\phi,\max}(T,L,A)$ where $f_{\phi,\max}$ is a tree-separable cost function defined as $\phi_{T,L,r}(x) = \max(\rho(T,L,r),x)$, with $\rho(T,L,r)=\max_{(L_u,L_v) \in Z, L_u=(K_1,K_2,K_3)}|K_3|$.
\end{definition}
The above function is tree-separable since $\phi_{T,L,r}$ and $\max$ satisfy the properties in Definition~\ref{def:treesep} and because $Z$ (and consequently $\phi$) depends only on $T$, $L$, $r$ and not on the rest of $A$.
This metric accurately computes the maximum buffer dimension passed through the root loop nest ($\rho(T,L)$), since the size of any buffer used in the fully fused loop nest tree is determined by the indices not yet iterated over (Equation~\ref{eqn:intermediate_indices}), namely those in $K_3$.
Further, since $\oplus$ is a max operator, the maximum buffer dimension needed within any inner loops is also considered by $f$ in a recursive manner.
This model can be modified to account for buffer size instead of dimension, by changing $r(A)$ to be the product of the dimensions of the indices in $K_3$.


\subsubsection{Total Number of Cache Misses}
\label{ssec:cost_function_mcs}
\hfill\\
To compute cost as the total number of cache misses for a given contraction path, we consider a simple cache model where the cache can hold $N$ subtensors of size $I^D$, where $I$ is the tensor dimension size and $N<I$.
For example, if $D=1$ and if the same column or row of a matrix is accessed consecutively, we assume the column or row is kept in cache.
We then model the number of cache misses incurred within each loop, by taking into any misses in contained (inner) loops and counting the number of tensors (inputs and outputs/intermediates computed) that are indexed by the loop index of this loop and still have at least $D$ other indices that need to be iterated over.
For each such tensor, at least $I^D$ distinct data from this tensor is loaded in each iteration of the loop, which incurs 1 cache miss.
Note that each cache miss in this model is associated with moving $I^D$ tensor data between memory and cache.
\begin{definition}[Cost Function for Total Number of Cache Misses]
Consider a fully fused loop nest tree $\mathcal{F}(A)$ for loop order $A$ with contraction path $(T,L)$.
Given a cache of size $I^D$, the number of cache misses is modeled by $f_{\phi,+}(T,L,A)$, where $f_{\phi,+}$ is a tree-separable cost function 
defined using $\phi_{T,L,r}(x) = I(r)(\tau(T,L,r)+x)$, where $I(r)$ is the dimension of the root index $r$ 
and
\begin{alignat*}{2}
\tau(T,L,r) =& |&&S|, \\
S =& \{&&v :  v\in (v_1,v_2,v_3) = L_u, \forall L_u \in L, \\
& &&  \text{ s.t. } r \in v \text{ and } |v| > D\}.
\end{alignat*}
\end{definition}
Again, it is easy to check that the defined cost function is tree-separable by properties of $\phi_{T,L,r}$ and $+$.
The cost function accurately captures the proposed cache miss model by multiplying the number of cache misses incurred in any loop iteration or its sub-loops by the number of loop iterations.
This model can be extended to consider other cache sizes, sparsity, multiple levels of cache, and cache line size.

\if showStriked
\subsubsection{Total Number of Cache Misses}
\label{ssec:cost_function_mcs}
\hfill\\
To compute cost as the total number of cache misses for a given contraction path, we consider a simple cache model where the cache can hold $N$ subtensors of size $I^D$, where $I$ is the tensor dimension size and $N<I$.
For example, if $D=1$ and if the same column or row of a matrix is accessed consecutively, we assume the column or row is kept in cache.
We then model the number of cache misses incurred within each loop, by taking into any misses in contained (inner) loops and counting the number of tensors (inputs and outputs/intermediates computed) that are indexed by the loop index of this loop and still have at least $D$ other indices that need to be iterated over.
For each such tensor, at least $I^D$ distinct data from this tensor is loaded in each iteration of the loop, which incurs 1 cache miss.
Note that each cache miss in this model is associated with moving $I^D$ tensor data between memory and cache.
\begin{definition}[Cost Function for Total Number of Cache Misses]
Consider a fully fused loop nest tree $\mathcal{F}(A)$ for loop order $A$ with contraction path $(T,L)$.
Given a cache of size $I^D$, the number of cache misses is modeled by $f_{\phi,+}(T,L,A)$, where $f_{\phi,+}$ is a tree-separable cost function 
defined using $\phi_{T,L,r}(x) = I(r)(\tau(T,L,r)+x)$, where $I(r)$ is the dimension of the root index $r$ 
and
\begin{alignat*}{2}
\tau(T,L,r) =& |&&S|, \\
S =& \{&&v :  v\in (v_1,v_2,v_3) = L_u, \forall L_u \in L, \\
& &&  \text{ s.t. } r \in v \text{ and } |v| > D\}.
\end{alignat*}
\end{definition}
Again, it is easy to check that the defined cost function is tree-separable by properties of $\phi_{T,L,r}$ and $+$.
The cost function accurately captures the proposed cache miss model by multiplying the number of cache misses incurred in any loop iteration or its sub-loops by the number of loop iterations.
This model can be extended to consider other cache sizes, sparsity, multiple levels of cache, and cache line size.
\fi

\begin{algorithm}
{\small
\caption{Algorithm to find cost-optimal loop order for terms in a given contraction path}
\label{alg:algo_recursive}
\begin{algorithmic}[1]
\Statex \textbf{Global Input:} 
Loop nest cost function $f$ specified for contraction path $(T,L)$ via parameterized scalar function $\phi$ and binary operator $\oplus$.
\Statex \textbf{Input:}
A contraction path $(T,L)$, with $L=(L_1,\dots,L_N)$, where each $L_i$ is a 3-tuple of index sets and $T$ is a binary contraction tree.
\Statex \textbf{Output:}
Two loop orders, $A$ and $B$, for $(T,L)$, so that $A$ has minimal cost ($f_{\phi,\oplus}(T,L,A)$) among all loop orders for $(T,L)$ and $B$ has minimal cost among all loop orders for $(T,L)$ that yield a loop nest tree $\mathcal{F}(B)$ with a different root than $\mathcal{F}(A)$.
\Procedure{\texttt{ORDER}}{$T$, $L$}
\State 
$\delta_{A} \gets \infty$;
$\delta_{B} \gets \infty$;
$A \gets \emptyset$; $B \gets \emptyset$
\If {$L = \emptyset$}
    \State \Return $(\emptyset,\emptyset)$
\EndIf
\If {$L[1] = \emptyset$}
    \State \Return \texttt{ORDER}($T\setminus L_1 , L\setminus L_1, \phi_{T\setminus L_1, L\setminus L_1}$)
\EndIf

\State $(u,v,w) = L_1$
\For {$q \in  u\cup v \cup w$}
  \State $\delta_{C} \gets \infty$; $C \gets \emptyset$
  \State $k \gets \max\limits_{k\in {1,\ldots, N}, \text{ s.t. } q\in L_1,\cdots, q\in L_k} k$ 
  \For {$s \gets 1 \text{ to } k$}
    \State Let $(T^{(X)},L^{(X)})$ be the contraction path restricted to the terms $L_1,\ldots, L_s$ with index $q$ removed.
    \State Let $(T^{(Y)},L^{(Y)})$ be the contraction path restricted to the terms $L_{s+1},\ldots, L_N$.
    \State $(A^{(X)}, \star) \gets      \texttt{ORDER}(T^{(X)}, L^{(X)})$
    \State $(\bar{A}^{(Y)},\bar{B}^{(Y)}) \gets \texttt{ORDER}(T^{(Y)}, L^{(Y)})$
    \LComment{If $Y$ tree has $q$ as root index, the resulting tree would be treated as not fully fused, so take second best tree.}
    \If {$\bar{A}^{(Y)}_{1}[1] = q$}
        \State $A^{(Y)} \gets \bar{B}^{(Y)}$
    \Else
        \State $A^{(Y)} \gets \bar{A}^{(Y)}$
    \EndIf
    \LComment{Compute cost of loop order.}
    \State $\delta \gets \phi_{T,L,q}\Big(f_{\phi,\oplus}(T^{(X)},L^{(X)},A^{(X)})\Big) \oplus f_{\phi,\oplus}(T^{(Y)},L^{(Y)},A^{(Y)})$ \label{li:alg_cost}
    \LComment{Update lowest cost loop orders}
    \If {$\delta < \delta_C$}
        \State $C \gets [[q,A_1^{(X)}],\ldots [q,A_s^{(X)}],A^{(Y)}]$
        \State $\delta_C \gets \delta$
    \EndIf
    \EndFor
    \If {$\delta_C < \delta_A$}
        \State $\delta_B \gets \delta_A$; $B\gets A$; $\delta_A \gets \delta_C$; $A \gets C$
    \ElsIf {$\delta_C < \delta_B$}
        \State $\delta_B \gets \delta_C$; $B\gets C$
    \EndIf
\EndFor
\State \Return $(A,B)$
\EndProcedure
\end{algorithmic}
}
\end{algorithm}

Algorithm~\ref{alg:algo_recursive} provides a fast search algorithm to find a cost optimal order for tree-separable cost functions.
In the pseudocode of the algorithm, for brevity, we use notation such as $T\setminus L_1$ to denote the tree obtained by removing the vertex in the contraction tree $T$ associated with the contraction term $L_1$.
We also use $[x,Y]$ to describe an item or list $x$ being prepended to list $Y$.

We now provide a proof of correctness and show how the subproblems of Algorithm~\ref{alg:algo_recursive} can be memoized to reduce its complexity.
For both, it is helpful to enumerate the subproblems (calls to function \texttt{ORDER}) in terms of 
\begin{enumerate}
\item the subsequence of terms included in the subproblem (size of $T$ and $L$),
\item the set of indices excluded from the terms (already iterated over), we refer to this set as $S$.
\end{enumerate}
We use induction on the size of these subproblems to prove correctness.

\begin{theorem}[Proof of Correctness of Algorithm~\ref{alg:algo_recursive}]
Consider a contraction path $(T,L)$ and a tree-separable cost function $f$ specified by $\phi_{T,L}$ and $\oplus$.
\texttt{ORDER}($T, L, \phi_{T,L,r}$) (Algorithm \ref{alg:algo_recursive}) returns two loop orders, $A$ and $B$, for $(T,L)$, so that $A$ has minimal cost ($f_{\phi,\oplus}(T,L,A)$) among all loop orders for $(T,L)$ and $B$ has minimal cost among all loop orders for $(T,L)$ that yield a loop nest tree $\mathcal{F}(B)$ with a different root than $\mathcal{F}(A)$.
\end{theorem}
\begin{proof}
We prove the theorem statement by induction on the size of $L$.
If there are no indices/terms remaining ($L=\emptyset$), only the null order is valid.
By inductive hypothesis, we assume the theorem statement holds
for any subsequence of terms in $L$ and the associated part of $T$ with any subset of indices removed from all terms in $L$ (the set of indices already iterated over contains $S$).
If the theorem statement does not hold, there must exist some order $A'$ for $(T,L)$ with $f_{\phi,\oplus}(T,L,A')<f_{\phi,\oplus}(T,L,A)$.
Let $r$ be the root of the first tree in $\mathcal{F}(A')$, $B^{(1)}$ be the first tree in the forest $\mathcal{F}(A')$, and $B^{(2)}$ be the remainder of the forest, with
and $(T^{(1)},L^{(1)})$ and $(T^{(2)},L^{(2)})$ being the associated contraction paths.
Since $f_{\phi,\oplus}$ is separable, we have that
\begin{align*}
f_{\phi,\oplus}(T,L,A')  
=&\phi_{T,L,r}(f_{\phi,\oplus}(T^{(1)},L^{(1)},B^{(1)})) \\
&\oplus f_{\phi,\oplus}(T^{(2)},L^{(2)},B^{(2)}).
\end{align*}
Since $(T^{(1)},L^{(1)})$ and $(T^{(2)},L^{(2)})$ are contained and smaller (as defined in our inductive hypothesis) than $(T,L)$, Algorithm~\ref{alg:algo_recursive}, when considering root vertex $r$, would return the minimal cost loop order for both subproblems.
Further, the cost of $A'$ would be computed correctly on line~\ref{li:alg_cost} of the Algorithm.
Since the algorithm instead found $A$ to have a lower cost, we have derived a contradiction.
Given optimality of $A$, its trivial to check that the given optimality condition for $B$ is maintained.
\end{proof}

\if showStriked
For both, it is helpful to enumerate the subproblems (calls to function \texttt{ORDER}) in terms of 
\begin{enumerate}
\item the subsequence of terms included in the subproblem (size of $T$ and $L$),
\item the set of indices excluded from the terms (already iterated over), we refer to this set as $S$.
\end{enumerate}
We now use induction on the size of these subproblems to prove correctness.

\begin{theorem}[Proof of Correctness of Algorithm~\ref{alg:algo_recursive}]
Consider a contraction path $(T,L)$ and a tree-separable cost function $f$ specified by $\phi_{T,L}$ and $\oplus$.
\texttt{ORDER}($T, L, \phi_{T,L,r}$) (Algorithm \ref{alg:algo_recursive}) returns two loop orders, $A$ and $B$, for $(T,L)$, so that $A$ has minimal cost ($f_{\phi,\oplus}(T,L,A)$) among all loop orders for $(T,L)$ and $B$ has minimal cost among all loop orders for $(T,L)$ that yield a loop nest tree $\mathcal{F}(B)$ with a different root than $\mathcal{F}(A)$.
\end{theorem}
\begin{proof}
We prove the theorem statement by induction on the size of $L$.
If there are no indices/terms remaining ($L=\emptyset$), only the null order is valid.
By inductive hypothesis, we assume the theorem statement holds
for any subsequence of terms in $L$ and the associated part of $T$ with any subset of indices removed from all terms in $L$ (the set of indices already iterated over contains $S$).
If the theorem statement does not hold, there must exist some order $A'$ for $(T,L)$ with $f_{\phi,\oplus}(T,L,A')<f_{\phi,\oplus}(T,L,A)$.
Let $r$ be the root of the first tree in $\mathcal{F}(A')$, $B^{(1)}$ be the first tree in the forest $\mathcal{F}(A')$, and $B^{(2)}$ be the remainder of the forest, with
and $(T^{(1)},L^{(1)})$ and $(T^{(2)},L^{(2)})$ being the associated contraction paths.
Since $f_{\phi,\oplus}$ is separable, we have that
\begin{align*}
f_{\phi,\oplus}(T,L,A')  
=&\phi_{T,L,r}(f_{\phi,\oplus}(T^{(1)},L^{(1)},B^{(1)})) \\
&\oplus f_{\phi,\oplus}(T^{(2)},L^{(2)},B^{(2)}).
\end{align*}
Since $(T^{(1)},L^{(1)})$ and $(T^{(2)},L^{(2)})$ are contained and smaller (as defined in our inductive hypothesis) than $(T,L)$, Algorithm~\ref{alg:algo_recursive}, when considering root vertex $r$, would return the minimal cost loop order for both subproblems.
Further, the cost of $A'$ would be computed correctly on line~\ref{li:alg_cost} of the Algorithm.
Since the algorithm instead found $A$ to have a lower cost, we have derived a contradiction.
Given optimality of $A$, its trivial to check that the given optimality condition for $B$ is maintained.
\end{proof}
\fi
We now consider the execution cost of Algorithm~\ref{alg:algo_recursive}, with the cost of each subproblem memoized.
For $N$ ordered terms and $m$ total indices, there are $O((m!)^N)$ loop orders (loop nests).
Algorithm~\ref{alg:algo_recursive} needs to consider all subsequences of the $N$ terms and all subsets of the $m$ indices, yielding $O(N^22^m)$ subproblems.
Each subproblem considers all choices of root index and prefixes of terms that contain that index to iterate over.
Thus the cost per subproblem is $O(mN)$ and the overall complexity of the algorithm is $O(N^32^mm)$.
\externaldocument{algorithm}
\vspace{-0.05in}
\section{SpTTN-Cyclops Framework}

We build a runtime framework for SpTTN kernels, which searches for cost-optimal loop nests using the methodology/algorithm introduced in Section~\ref{sec:algorithm} and executes the resulting loop nests.
Specifically, the framework first considers all contraction paths with optimal asymptotic complexity.
For each contraction path, we restrict loop orders to those in which the indices of the sparse tensor are iterated over in the order in which they are stored in the CSF tree.
We select the minimum cost loop nest among these using Algorithm~\ref{alg:algo_recursive}.
If the framework cannot find a loop nest that fits within the constraints set by the cost model, it iterates over the contraction paths with suboptimal asymptotic complexity until it finds a loop nest that adheres to the constraints.
While the framework may use different cost functions and employ autotuning, in the experiments, we use a tree-decomposable cost metric that selects the loop nest with the maximum number of independent dense loops with bounded buffer dimension.
This choice is made to use BLAS kernels as much as possible while maintaining a bounded amount of storage.

\begin{figure*}
\begin{subfigure}{0.33\textwidth}
\includegraphics[scale=.45]{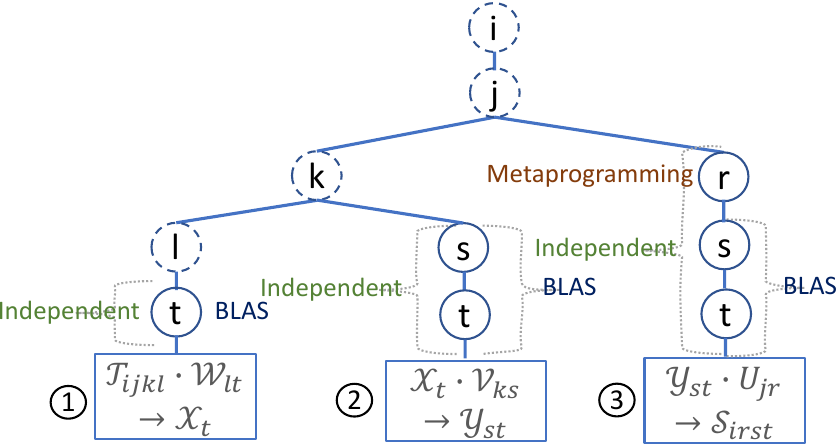}
\end{subfigure}
\qquad
\begin{subfigure}{0.56\textwidth}
\begin{lstlisting}[belowskip=0em]
T_csf = CSF(${\tsr{T}\_ijkl}$)
for (i,T_i) in T_csf:
  for (j,T_ij) in T_i:
    $\tsr{Y}$ = 0 // reset intermediate tensor
    for (k,T_ijk) in T_ij:
      $\tsr{X}$ = 0 // reset intermediate tensor
      for l in T_ijk:
        |{\textcolor{lred}{\textbf{xAXPY}}}|(T, t_ijkl, $\tsr{W}$[l,:], 1, $\tsr{X}$, 1)
      |{\textcolor{lred}{\textbf{xGER}}|(T, S, 1, $\tsr{X}$, 1, $\tsr{V}$[k,:], 1, $\tsr{Y}$, T)}
    for r in range(R): // dense loop
      |\textcolor{lred}{\textbf{xAXPY}}|((S * T), $\tsr{Y}$, $\tsr{U}$[j,r], 1, $\tsr{S}$[i,r,:,:], 1)
\end{lstlisting}
\end{subfigure}
\caption{Loop nest for an order 4 TTMc kernel. 
Loop $r$ of contraction \circled{3} is not via recursion but is generated as a loop by metaprogramming.
Contractions \circled{1} and \circled{3} are offloaded to BLAS-1, and contraction \circled{2} is offloaded to a BLAS-2 kernel.}
\label{fig:ttmc_order_4}
\end{figure*}

\vspace{-0.05in}
\subsection{Algorithm to Generate and Execute Loop Nests}
\label{ssec:runtime_algo}
Given a fully fused loop nest tree, in Algorithm \ref{alg:recursive} we present a runtime algorithm to generate loop nests and execute the contractions. 
We represent the tree with a sequence of terms (leaves) and a list per term representing the loop order (vertices). 
This representation is sufficient for the algorithm to infer the structure of a fully fused loop nest tree. 
We use Algorithm \ref{alg:recursive} in two stages. In the first stage, we preprocess the fully fused loop nest tree and add hooks to (1) generate nested loops for the dense indices using metaprogramming, (2) identify independent dense loops that can be offloaded to BLAS like kernels. We also allocate memory for the intermediate tensors in this stage.
In the second stage, we compute the kernel by executing the preprocessed fully fused loop nest tree. We check for hooks in Line 2 and offload the computation accordingly. 

\begin{algorithm}
  \caption{Algorithm to generate loop nests}
  \label{alg:recursive}
  \begin{algorithmic}[1]
  \Statex \textbf{Input:} Sequence of \textit{terms} that represent the
  \textit{contraction path}. Each term is a set 
  of three tensors, inp1, inp2 and op. 
  \Statex \textbf{Input:} Depth \textit{initially set to 0}.
  \Statex \textbf{Output:} Loop nest to compute the given kernel. 
  \Procedure{\texttt{LOOP\_NEST}}{sequence\_of\_terms,\\ \hspace*{9.6em} depth}
  \If {$\text{depth} = |\text{sequence\_of\_terms}[0].\text{idx\_order}|$}
    \State {$t \gets \text{sequence\_of\_terms}[0]$}
    \State $\text{contract}(t.\text{inp1}, t.\text{inp2}, t.\text{op})$
  \EndIf
  \State $\text{idx} \gets \text{sequence\_of\_terms}[0].\text{idx\_order[depth]}$
  \State $\text{buf\_terms} \gets \emptyset$
  \For {$c \in \text{sequence\_of\_terms}$}
      \If {$\text{idx} = c.\text{idx\_order[depth]}$}
        \State $\text{buf\_terms} \gets {\text{buf\_terms} \cup c}$
      \Else
        \If {$|\text{buf\_terms}| \ge 1$}
          \For {$i \gets 1,\text{|buf\_terms|}$}
            \State $b \gets \text{buf\_terms}[i]$
            \State $\text{reset} \gets \text{True}$
            \For {$j \gets i+1, \text{buf\_terms}$}
              \If {$b.\text{op} = \text{buf\_terms}[j].\text{inp1}$ or $b.\text{op} = \text{buf\_terms}[j].\text{inp2}$}
                \State $\text{reset} \gets \text{False}$
              \EndIf
            \EndFor
            \If {$\text{reset} = \text{True}$}
              \State $b.\text{op} \gets 0$
            \EndIf
          \EndFor
          \LComment{generate a loop for idx}
          \State LOOP\_NEST($\text{buf\_terms},\ \text{depth} + 1$)
        \EndIf
        \State $\text{buf\_terms} \gets \emptyset$
        \State $\text{idx} \gets c.\text{idx\_order[depth]}$
      \EndIf
  \EndFor
  \If {$|\text{buf\_terms}| \ge 1$}
    \LComment{generate a loop for idx}
      \State LOOP\_NEST($\text{buf\_terms},\ \text{depth} + 1$)
  \EndIf
  \EndProcedure
  \end{algorithmic}
  \end{algorithm}

\vspace{-0.05in}
\subsection{Data Distribution}
We leverage CTF's \cite{ctf} data distribution strategy, which uses a cyclic data layout on multidimensional processor grids to~\nopagebreak[0] achieve load balance and scalability for sparse tensor computations.
We continue to hold the main sparse tensor in the same layout for the entire duration of the execution. 
Each dimension of the tensor is distributed across the processor grid in a cyclic fashion.
We redistribute the dense tensors, including the output tensor (if it is dense), along the dimensions it shares with the sparse tensor. 
Let $\{i_{1},\ldots,i_{r}\}$ be the indices of a dense tensor $\tsr{D}$ with dimensions $I_{1} \times \ldots \times I_{r}$. 
Assume a single index of $\tsr{D}$, $i_k$, is shared with the sparse tensor. Let the processor grid be $P_1 \times \ldots \times P_n$ and assume $i_k$ is mapped to $P_j$. Then, $D$ is partially replicated so that all processors $q_1,\ldots,q_j$ with a fixed index $q_j$ own all elements of $\tsr{D}$, or which $i_k \equiv q_j \bmod P_j$.
Note that in tensor decomposition and completion algorithms these replicated dimensions are often relatively small. Each processor can now perform local kernel computation without any further data exchange.
After the computation we reduce the output tensor and redistribute it to its original mapping on the processor grid.

\externaldocument{algorithm}
\externaldocument{detailed_design}
\subsection{Example SpTTN Execution}
\label{ssec:example_spttn_ttmc_o4}
\if showStriked
\subsection{End-to-end Example of SpTTN Execution}
Consider the computation of an order 4 TTMc kernel, 
\[\tsr{S}(i,r,s,t) = \sum_{j,k,l} \tsr{T}(i,j,k,l) \cdot \tsr{U}(j,r) \cdot \tsr{V}(k,s) \cdot \tsr{W}(l,t).\]

\textbf{Autotuner: } The autotuner tunes over various ways to contract the tensors, 
$(\tsr{T}\cdot\tsr{U}\cdot\tsr{V}\cdot\tsr{W})\rightarrow\tsr{S}$, and the corresponding indices in the contraction terms, to pick an optimal contraction path and index order. 
An optimal contraction path in terms of asymptotic complexity is the one that has a maximum loop depth of 5.
The permutations of the index order are restricted to orders that have the sparse indices $i,j,k,l$ appear as they are stored in the CSF representation of $\tsr{T}$. 
The search space is further pruned by evaluating the schedules based on the cost models presented in Section \ref{sec:cost_optimal_loop_nests} as discussed earlier in this section. \\
\textbf{Preprocessor:} Let the contraction path of terms and their corresponding index order picked by the autotuner be 
$((\tsr{T}\cdot\tsr{W}\rightarrow\tsr{X}), (\tsr{X}\cdot\tsr{V}\rightarrow\tsr{Y}), (\tsr{Y}\cdot\tsr{U}\rightarrow\tsr{S}))$ and $((i,j,k,l,t),$ $(i,j,k,s,t), (i,j,r,s,t))$,
respectively. $\tsr{X}$ and $\tsr{Y}$ are intermediate tensors.
In Figure \ref{fig:ttmc_order_4}, we show the corresponding fully fused loop nest. 
We preprocess the terms to identify independent indices and generate tags to offload them to optimized dense kernels. 
The independent dense indices of the three terms are $(t), (s,t), (r,s,t)$, respectively.
The independent dense indices $\{t\}$ and $\{s,t\}$ of terms \circled{1} and \circled{2}, respectively, are tagged to use BLAS kernels. From the indices $(r,s,t)$ of term \circled{3}, the last two indices are tagged to use a BLAS kernel, and the first index, $r$, is tagged as a dense loop to be generated by metaprogramming. \\
\textbf{Intermediate tensors: } Next, we assign indices for the intermediate tensors $\tsr{X}$ and $\tsr{Y}$, and allocate memory for them. $\tsr{X}$ is an intermediate tensor between \circled{1} and \circled{2}. 
We compute $((i,j,k,l,t)\cap (i,j,k,s,t))\backslash(i,j,k)$ to get $(t)$ as the index of $\tsr{X}$. Similarly, we intersect $((i,j,k,s,t)\cap (i,j,r,s,t))\backslash(i,j)$ to get $(s,t)$ as indices of $\tsr{Y}$. \\
\textbf{Data distribution: } The sparse tensor is retained in the initial distribution. Tensors $\tsr{U}$, $\tsr{V}$, $\tsr{W}$ and the output tensor $\tsr{S}$ share an index each with the sparse tensor. Indices that are not shared are replicated on each processor i.e., $r\in\tsr{U}$, $s\in\tsr{V}$, $t\in\tsr{W}$ and $(r,s,t)\in\tsr{S}$ are replicated. 
The dimensions $I, J, K$ and $L$, which are large in size, maintain their original distribution on the processor grid. \\
\textbf{Runtime: } 
Finally, we use the recursive algorithm (Algorithm \ref{alg:recursive}) to generate the loop nests at runtime and check for an opportunity to offload the independent dense loops based on the hooks generated in the preprocessing step.
We show the final fully fused loop nest with the BLAS kernels in Figure \ref{fig:ttmc_order_4}.
After the computation, we perform reduction on the tensor $\tsr{S}$ and redistribute it back to its original mapping on the processor grid.
\fi

In Figure \ref{fig:ttmc_order_4}, we show a fully fused loop nest for the order 4 TTMc kernel,
$\tsr{S}(i,r,s,t) = \sum_{j,k,l} \tsr{T}(i,j,k,l) \cdot \tsr{U}(j,r) \cdot \tsr{V}(k,s) \cdot \tsr{W}(l,t)$.

\section{Related Work}
\textbf{General tensor algebra compilers:} 
TACO \cite{TACO} and COMET \cite{comet} consist of Domain Specific Language (DSL) compilers to generate kernels for both sparse and dense tensors. The default schedules of these frameworks are unfactorized and can be suboptimal for SpTTN kernels.


SparseLNR~\cite{SparseLNR} and ReACT~\cite{ReACT} extend TACO and COMET, respectively, with kernel distribution/fusion to support the factorize-and-fuse approach.
The contraction path and loop orders for these loop nests are user-specified.
Our main contribution is in fully enumerating the space of loop nests and finding a cost-optimal schedule automatically.
Furthermore, in our evaluation (in Section \ref{sec:evaluation}), we show that SpTTN-Cyclops outperforms SparseLNR by orders of magnitude. For example, across various input tensors considered, SpTTN-Cyclops outperforms SparseLNR by 1.3x to 3.4x and 4x to 110.5x on MTTKRP and TTMc kernels, respectively.\\ 
\conditionalPara{Mosaic~\cite{mosaic} is a recent work that proposes a sparse tensor algebra compiler that can bind tensor sub-expressions to external functions of other tensor algebra libraries and compilers, and provides an automated search mechanism to find all valid bindings. In our framework, we automatically bind the innermost loops to other optimized routines (like BLAS) at runtime.}
\textbf{Auto-scheduler:} 
Tensor Contraction Engine (TCE)~\cite{TCE} automatically generates sequence of tensor contractions that minimize intermediate tensor sizes. It primarily focuses on dense tensor operations that are common in quantum chemistry computations.
The dynamic programming approach in TCE~\cite{tce_dp,tce_dp_tiling} adopts a bottom-up approach i.e., to find an optimal loop structure, the subtrees of the loop nest tree are evaluated first and memoized. Subsequently, at the root node, various loop structures including the possibility of fusing the subtrees are evaluated to pick the optimal loop structure.
Furthermore, in TCE, the tree is partitioned into sub-problems by identifying a set of cut-points. There can be multiple cut-points at a given level. In SpTTN-Cyclops, at any given iteration, we split the problem into two sub-problems, i.e., only the first cut-point is considered, and the cost of the sub-problems is memoized. So a subproblem is a choice for the root index and prefixes of terms that contain that index to iterate over. This approach of SpTTN-Cyclops reduces the cost (for finding an optimal loop nest) when compared to choosing an index for each subtree at a given level and translates into better search complexity.

Protocolized Concrete Index Notation (CIN-P) \cite{cinp}, proposes an automated scheduler that  enumerates every schedule of minimum depth and relies on the kernel being small. 
CIN-P focuses solely on asymptotic costs and CIN-P for TACO discards schedules involving intermediate tensors of more than one dimension. 
SpTTN-Cyclops on the other hand tunes over both contraction path and loop orderings.
WACO~\cite{waco} co-optimizes the format and schedule of sparse tensor kernels using a sparse convolutional neural network to model and predict the runtime performance based on the sparsity patterns, formats, and schedules.
SparseAuto~\cite{dias2024sparseauto} prunes the search space of schedules for sparse tensor contractions based on both time and intermediate tensor memory requirements. It uses Satisfiability Module Theory (SMT) solvers to pick the smallest number of possible schedules based on user-defined constraints.
In CoNST~\cite{const_auto}, the authors use a constraint-based approach with a Z3 SMT solver to optimize schedules for sparse tensor contractions.

Inspector-executor models incorporated in the compiler transformation frameworks such as Sparse Polyhedral Framework (SPF)~\cite{spf1, spf2} enable optimization of sparse computations. 
In \cite{spf_tensor}, the authors extend SPF to generate optimized sparse tensor codes.
They focus on kernels that handle multiple sparse tensors and not SpTTN kernels. 
\\
\textbf{General distributed-memory frameworks:} DISTAL \cite{DISTAL} extends TACO to target distributed systems. SpDISTAL \cite{SpDISTAL} adopts single-node transformations of TACO and extends DISTAL
with new constructs for describing distributions of sparse tensors. 
SpDISTAL inherits the limitations of TACO in terms of finding an optimal code path for SpTTN kernels. Also, our framework provides automatic distributed memory parallelization without any user intervention.
Deinsum \cite{Deinsum} provides automatic distributed-memory parallelization of operations on dense tensors.
TiledArray~\cite{tiledarray, tiledarray_software} is a distributed-memory framework for block-sparse tensors.
\\
\textbf{Specialized library implementation for SpTTN kernels:}~SPLATT~\cite{SPLATT} provides an optimized implementation~\nopagebreak[0]~of MTTKRP on shared and distributed memory systems. GigaTensor \cite{GigaTensor} implements MTTKRP as a series of Hadamard products and uses the MapReduce paradigm. A parallel algorithm for TTMc which leverages multiple CSF representations is proposed in \cite{accelerate-tucker}.
Parallel Tensor Infrastructure (ParTI!) \cite{parti} is a library for sparse tensor operations (including MTTKRP) and tensor decompositions on multicore CPU and GPU architectures.
In \cite{parti_reorder}, as part of ParTI!, the authors propose techniques to reorder the sparse tensor to improve the performance of MTTKRP.

\externaldocument{background}
\externaldocument{algorithm}
\section{Evaluation}
\label{sec:evaluation}
\begin{figure}[h]
\centering
\includegraphics[scale=0.23]{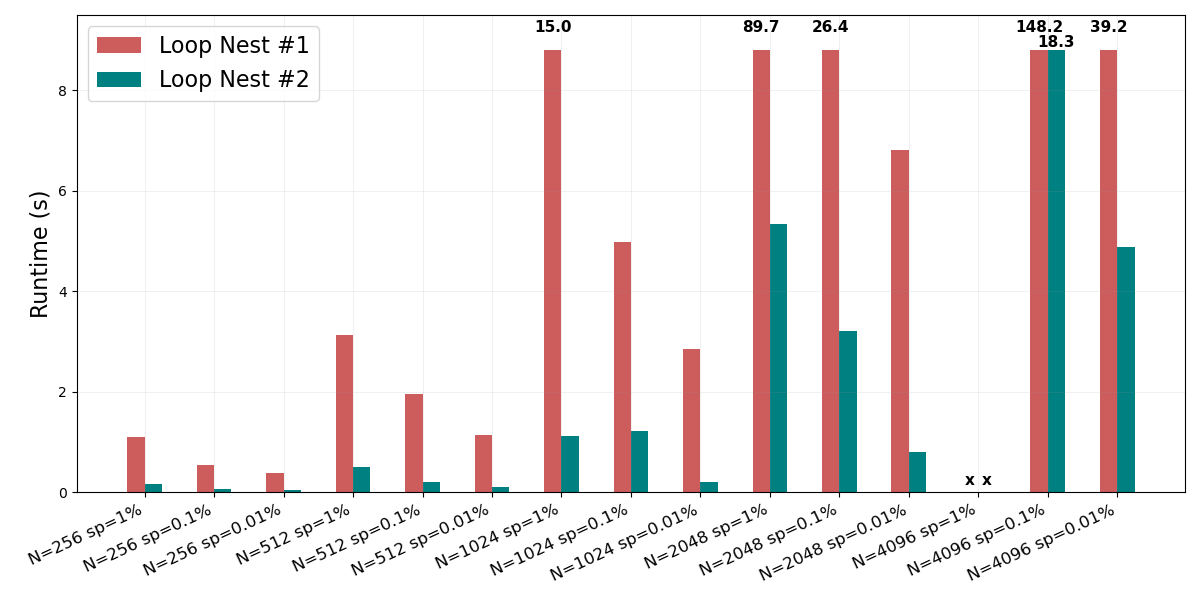}
\vspace{-0.1in}
\caption{Single thread performance of an order 3 all-mode TTMc contraction. Loop Nest \#1 has a bound of 1 and Loop Nest \#2 has a bound of 2 on the intermediate tensor dimension. $R=64$.}
\label{fig:ttmc_allm_2s}
\vspace{-0.2in}
\end{figure}

\begin{figure}[h]
\centering
\includegraphics[scale=0.23]{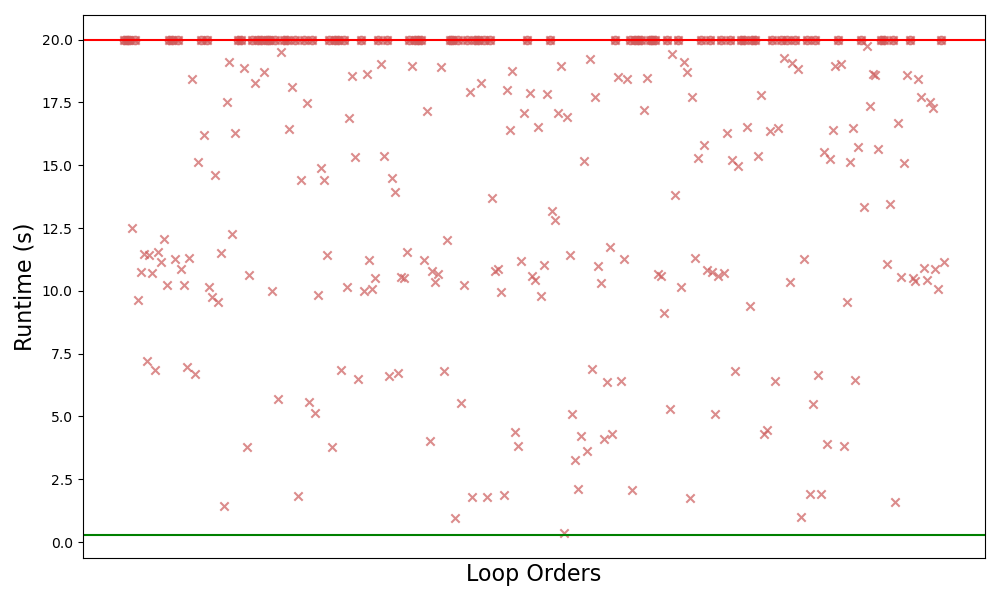}
\vspace{-0.1in}
\caption{Single thread performance of an order 3 all-mode TTMc contraction with $N=1024$, $R=32$, and sparsity at $0.1\%$ using randomly picked loop orders. The red line represents the cut-off and the green line represents the runtime of the loop order picked by SpTTN-Cyclops.}
\label{fig:ttmc_allm_ss}
\end{figure}

All results are collected on the Stampede2 supercomputer. Each node has an Intel Xeon Phi 7250 CPU (“Knights Landing”) with 68 cores, 96GB of DDR4 RAM, and 16GB of high-speed on-chip MCDRAM memory. 
Additionally, we also run our experiments on a single node equipped with an Intel Xeon Silver 4314 processor ("Ice Lake"), which features a 64KB L1 data cache per core, 1MB L2 cache per core, and a 24MB shared L3 cache. 
The results of these experiments are reported in Figures \ref{fig:ttmc_allm_2s} and \ref{fig:ttmc_allm_ss}.
In our distributed memory experiments, we use 64 MPI processes per node. We select a loop nest with the maximum independent dense loops by imposing a bound on the intermediate tensor dimension, maintaining it at two.
We compare SpTTN-Cyclops with TACO~\cite{TACO} (commit ID 2b8ece4), general sparse pairwise contraction with CTF~\cite{ctf} (v1.5.5, commit ID 36b1f6d), SPLATT~\cite{SPLATT} (v1.1.0, commit ID 6cb8628) and SparseLNR~\cite{SparseLNR} (branch dev-fuse, commit ID 8fafdd1).
We present results for single thread performance comparing with CTF, TACO, SparseLNR and SPLATT.
In SparseLNR, we try to use the optimal schedule derived from SpTTN-Cyclops, using its directives for kernel distribution and loop fusion. In TACO, we use the contraction path picked by SpTTN-Cyclops.
For distributed memory performance we compare against CTF and SPLATT. 

\begin{figure}[h]
\centering
\includegraphics[scale=0.38]{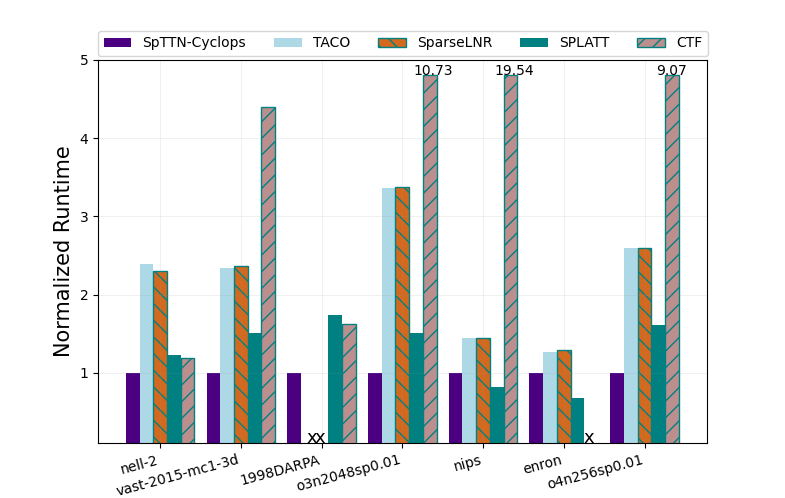}
\vspace{-0.1in}
\caption{Single thread performance of MTTKRP with $R=64$.}
\label{fig:single_node_mttkrp}
\end{figure}  

\begin{figure*}
\subfloat[TTMc]{\includegraphics[width=.34\textwidth]{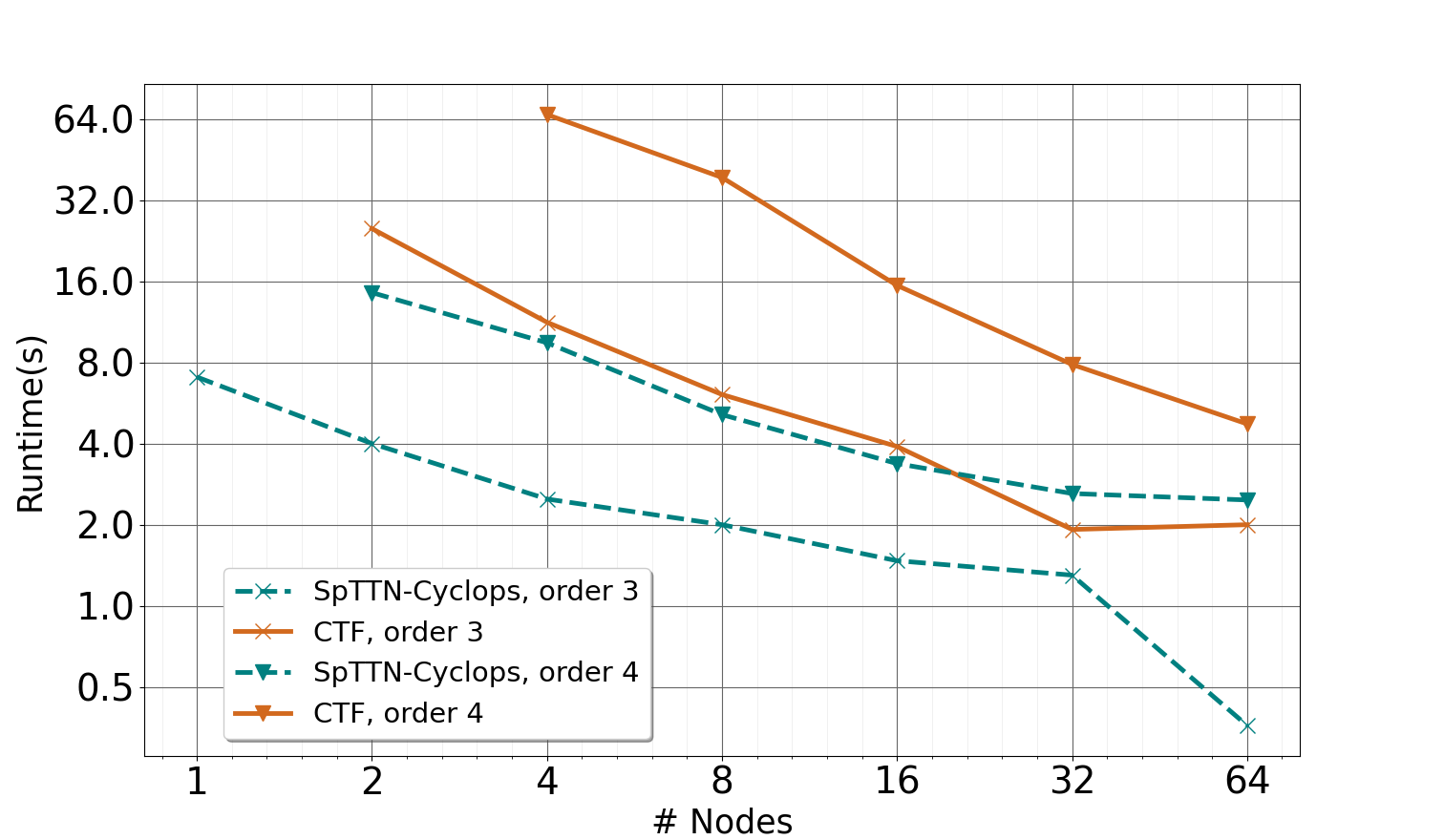}}\hskip -2ex
\subfloat[MTTKRP]{\includegraphics[width=.34\textwidth]{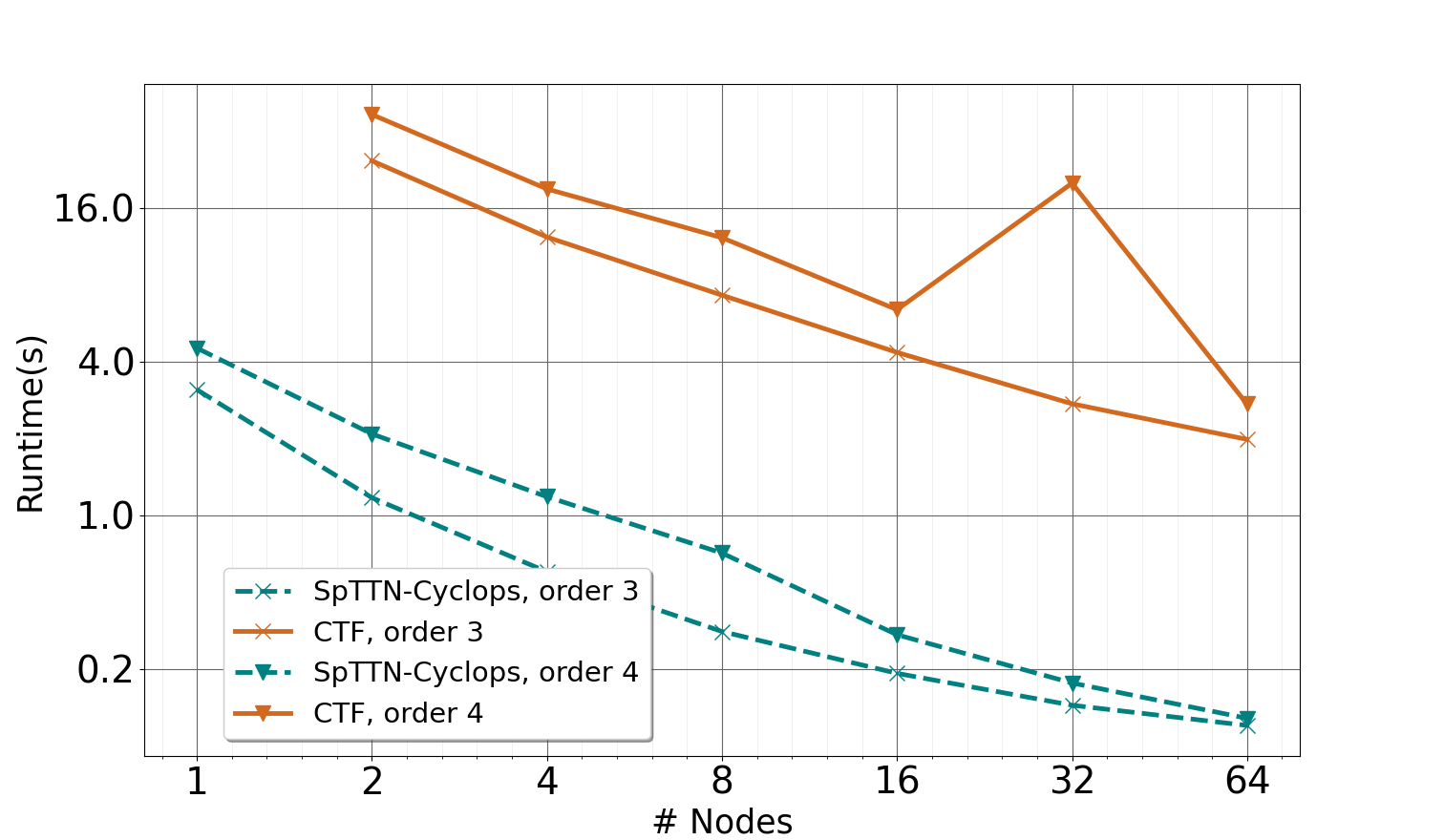}}\hskip -2ex
\subfloat[TTTP]{\includegraphics[width=.34\textwidth]{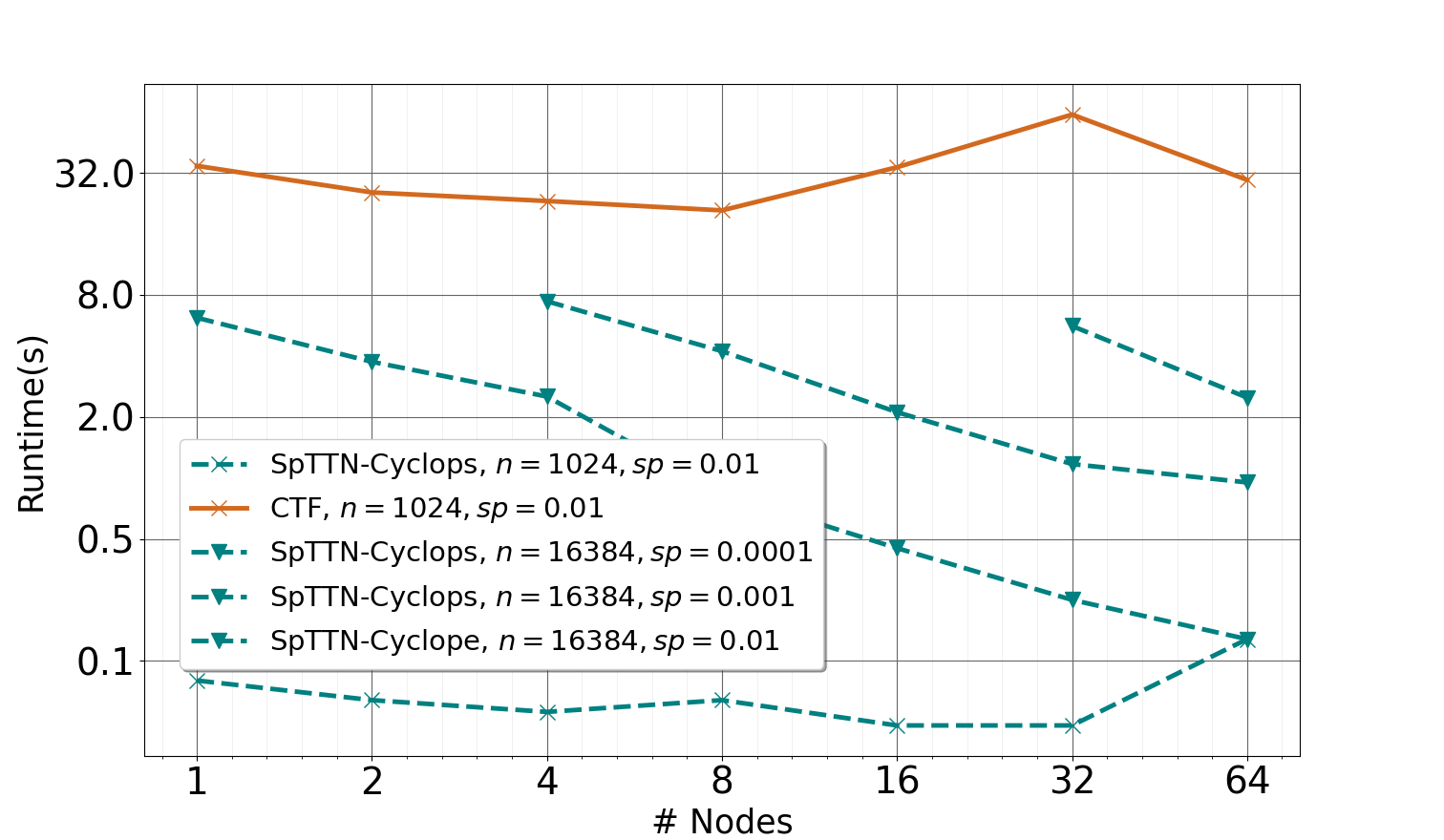}}\\
\vspace{-0.1in}
\caption{Strong scaling of kernels TTMc, MTTKRP and TTTP. 
The sparse tensor dimensions are identical across all modes.
TTMc and MTTKRP are computed on order 3 and order 4 tensors of $0.1\%$ sparsity. 
Their dimensions are set to $8192$ and $1024$, respectively.
TTTP is computed on order 3 tensors. $R=32$.} 
\label{fig:scaling}
\vspace*{-0.1in}
\end{figure*}



\if showStriked
\begin{figure}[h]
\centering
\includegraphics[width=.34\textwidth]{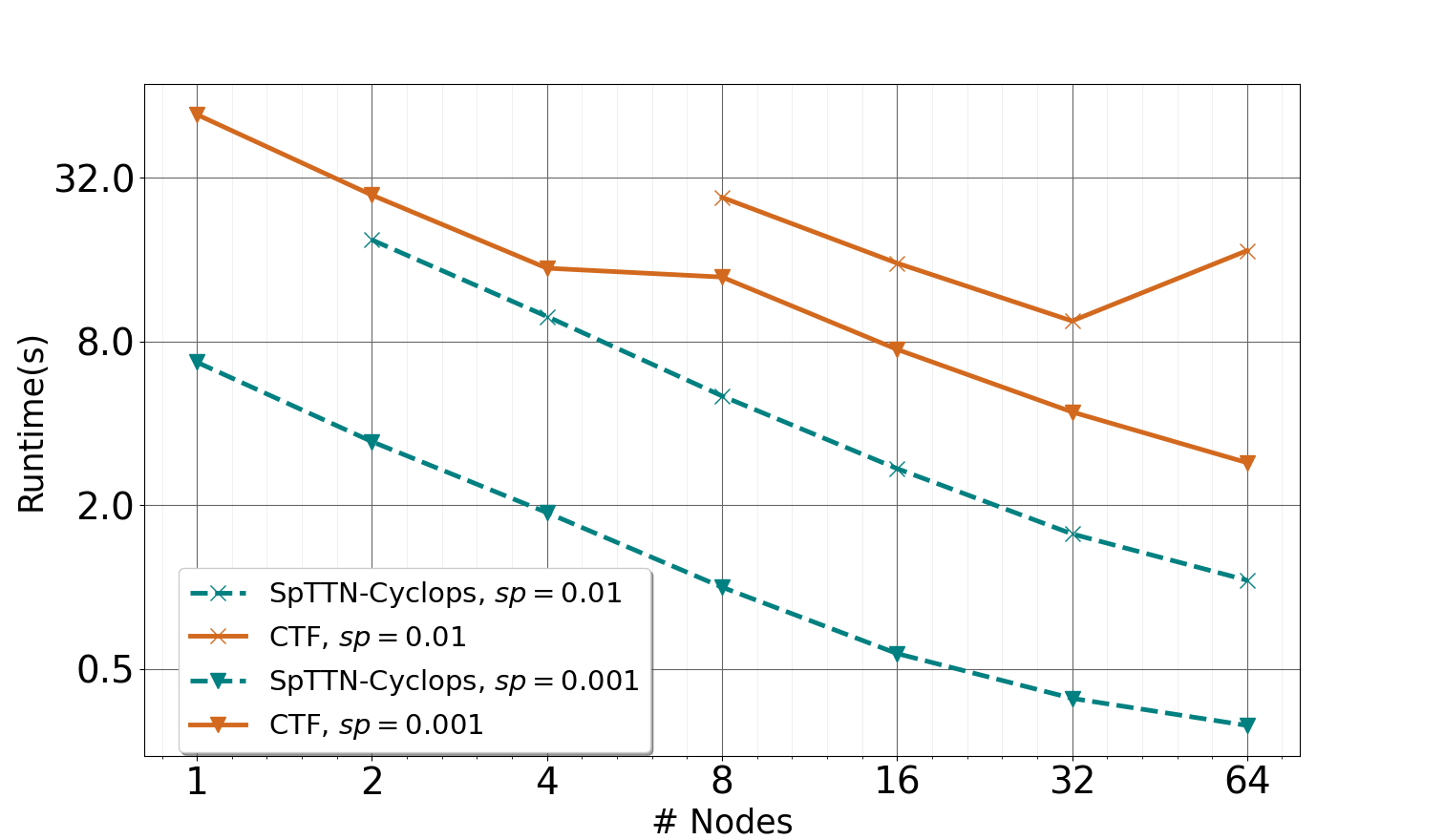}
\caption{Strong scaling of TTTc on order 6 tensors of sparsity $1\%$ and $0.1\%$. $N$ and $R$ are set to $80$ and $16$, respectively.}
\label{fig:scaling_k6}
\end{figure}  
\fi


\noindent \textbf{Datasets:} 
To evaluate the kernels, we use sparse tensors from Formidable Repository of Open Sparse Tensors and Tools (FROSTT) \cite{frostt} and 1998 DARPA Intrusion Detection Evaluation \cite{hanten2}. 
For further analysis, we generate random sparse tensors with various dimensions and sparsities. 
The dense tensors are populated with random data / nonzero positions. 
If a tensor has identical dimensions, $N$ is used to represent size of the dimensions. 
A dense tensor shares some of its indices with the sparse tensor. 
The dimensions of the non-shared indices are denoted using $R$. 
\\
\textbf{MTTKRP:} 
In Figure \ref{fig:single_node_mttkrp}, we compare the single thread performance of SpTTN-Cyclops to that of the other frameworks.
One of the optimal schedules to compute an order 3 MTTKRP in Equation \ref{eqn:mttkrp} is to have a loop nest that partially contracts $\tsr{T}$ with $\tsr{U}$, and then with $\tsr{V}$. This reduces the number of operations when compared to an unfactorized approach of TACO. SpTTN-Cyclops and SPLATT implement this factorize-and-fuse approach.
We observe speedups of $1.3$x to $3.4$x when compared to TACO. SparseLNR fails to fuse loops for this kernel and has similar performance to TACO. 
SpTTN-Cyclops achieves speedups of $1.5$x to $1.7$x, and slowdowns of $0.8$x and $0.7$x when compared to SPLATT.
CTF performs poorly when computing MTTKRP across all tensors.
We also conduct strong scaling experiments for MTTKRP on the \emph{nell-2} tensor. Despite being generic, our approach has performance close to SPLATT, a library optimized for a specific kernel.\\
\textbf{TTMc:} 
We observe substantial speedups over TACO. Since we factorize the kernel into pairwise contractions and then fuse loops in SpTTN-Cyclops, there is an asymptotic reduction in computation complexity which translates to these observed speedups.
For an order 3 TTMc kernel (Equation \ref{eqn:ttmc}), SparseLNR generates a schedule that contracts $\tsr{T}$ with $\tsr{U}$, and the result with $\tsr{V}$. Only index $i$ is fused across the two pairwise contractions, and requires an intermediate tensor of $K\times R$ dimension. If the input tensor expression is $\tsr{S}(i,r,s) = \sum_{j,k} \tsr{T}(i,j,k) \cdot \tsr{V}(k,s) \cdot \tsr{U}(j,r)$, i.e., the position of $\tsr{U}$ and $\tsr{V}$ are interchanged in the expression, then SparseLNR defaults to the unfactorized approach of TACO. SpTTN-Cyclops generates a schedule that contracts $\tsr{T}$ with $\tsr{V}$, and the result with $\tsr{U}$. Indices $i$ and $j$ are fused, and the intermediate tensor dimension is $S$ (Listing \ref{lst:ttmc_fusion}). 

For an order 4 TTMc kernel (Section \ref{ssec:example_spttn_ttmc_o4}), SparseLNR generates a schedule that contracts tensors $\tsr{T}$, $\tsr{U}$ and $\tsr{V}$ all-at-once, and the resulting intermediate tensor with $\tsr{W}$. The intermediate tensor dimension is $L\times R\times S$. Only index $i$ is fused across these two contractions. The maximum loop depth is six. SpTTN-Cyclops generates an asymptotically optimal schedule as shown in Figure \ref{fig:ttmc_order_4}, which has a maximum loop depth of five. 

We are able to run TTMc with TACO and SparseLNR only on two of the considered tensors, \emph{nell-2} and \emph{vast-3d}. On \emph{nell-2}, we observe a speedup of $29.3$x and $110.5$x over TACO and SparseLNR, respectively. Similarly, in \emph{vast-3d}, we observe a speedup of $125.9$x and $4$x.
We observe speedups over CTF in the range of $0.8$x to $12.6$x for the tensors considered (we are unable to run TTMc with CTF on \emph{enron} and \emph{nell-2} tensors).
On the \emph{nips} tensor where the combination of the imbalanced dimensions of the tensor and the specific value of $R$ does not benefit the fused approach of SpTTN-Cyclops, we see a slowdown of $0.8$x. 
We are unable to execute TTMc on \emph{darpa} using any of the approaches including SpTTN-Cyclops because of the larger memory footprint requirement for the contraction that cannot be accommodated on a single node.
In Figures \ref{fig:scaling}(a) and \ref{fig:scaling}(b), we present strong scaling results for MTTKRP and TTMc, respectively. SpTTN-Cyclops outperforms CTF on all node counts, and shows good scaling for both the kernels.\\
\textbf{TTTP:}
We present strong scaling results for TTTP in Figure \ref{fig:scaling}(c). The single node performance of SpTTN-Cyclops over CTF is substantial with over $340$x speedup. 
We observe good scaling for all the considered tensors. \\
\textbf{TTTc:}
In our strong scaling analysis of TTTc, we evaluate two tensors of dimension $80$ ($R=16$) and sparsity at $1\%$ and $0.1\%$.
In both, SpTTN-Cyclops achieves good scaling. SparseLNR generates a default TACO schedule for this kernel. We are unable to run TTTc implementation in TACO and SparseLNR on these kernels with the considered dimensions. However, we generated a smaller tensor with dimensions $N=40$ and sparsity at $0.1\%$. SpTTN-Cyclops achieves a speedup of $534$x over TACO 
on it. \\ 
\textbf{Impact of intermediate tensor dimension:}
Consider an order 3 all-mode TTMc kernel, $\tsr{S}(r,s,t) = \sum_{i,j,k} \tsr{T}(i,j,k) \cdot \tsr{U}(i,r) \cdot \tsr{V}(j,s) \cdot \tsr{W}(k,t)$ (all sparse indices are contracted). The contraction path chosen by SpTTN-Cyclops is $((\tsr{T}_{ijk}\cdot \tsr{W}_{kt}\rightarrow\tsr{X}_{ijt}),(\tsr{X}_{ijt}\cdot \tsr{V}_{js}\rightarrow\tsr{Y}_{ist}),(\tsr{Y}_{ist}\cdot \tsr{U}_{ir}\rightarrow \tsr{S}_{rst}))$.
For the chosen contraction path, if we consider a bound of two on the intermediate tensor dimension, the loop nest generated by SpTTN-Cyclops, $((i,j,k,t),(i,j,s,t),(i,r,s,t))$, has intermediate tensors $\tsr{X}$ of size $T$ and $\tsr{Y}$ of size $S\times T$.
For the same contraction path, if we consider a bound of one on the intermediate tensor dimension, the loop nest generated,
$((i,t,j,k),(i,t,j,s),(i,t,r,s))$, has intermediate tensors $\tsr{X}$ of size $1$ (scalar) and $\tsr{Y}$ of size $S$.
In Figure~\ref{fig:ttmc_allm_2s}, we show the single thread performance of the two loop nests generated by SpTTN-Cyclops for the order 3 all-mode TTMc kernel. We observe that the loop nest with intermediate tensors of size $T$ and $S\times T$ performs better than the loop nest with intermediate tensors of size $1$ and $S$, despite having a larger memory footprint.
The contractions in Loop Nest \#2 are offloaded to two \texttt{xAXPY} (BLAS-1) (manually-implemented) and one \texttt{xGER} (BLAS-2) kernels. 
Loop Nest \#1, on the other hand, employs an innermost sparse loop to compute the intermediate tensor $\tsr{X}$. 
Consequently, only two BLAS kernels are used in this loop nest: one for computing $\tsr{Y}$ and the other for $\tsr{S}$. \\ 
\textbf{Impact of loop order:} 
For the order 3 all-mode TTMc kernel and the contraction path chosen by SpTTN-Cyclops, we randomly select $25\%$ of all possible loop orders that have the sparse indices iterated over in the order in which they are stored in the CSF tree. In Figure~\ref{fig:ttmc_allm_ss}, we show the single thread performance of these randomly picked loop orders. In the loop order picked by SpTTN-Cyclops the intermediate tensors are within a considerable memory bound and also allows for the maximum use of BLAS kernels.

\vspace{-0.05in}
\section{Conclusion and Future Work}
Favorable performance of SpTTN-Cyclops in comparison to other general tensor contraction libraries, as well as comparisons to specialized codes, demonstrate that implementation of high-performance SpTTN kernels of interest to tensor decomposition and completion can be effectively automated.
As opposed to prior frameworks for sparse tensor contractions, by restricting consideration to a single sparsity pattern and dense buffers, we are able to enumerate and efficiently find the minimum cost SpTTN loop nest.
At the same time, the resulting implementations are practical, as they may be accelerated by standard BLAS libraries, and match the structure of existing optimized codes specialized to particular SpTTN contractions.
Our framework and evaluation of SpTTN kernels can be extended in several ways. For example, the search space can be extended to include partially-fused loop nests, which may offer additional parallelism.
\begin{acks}
This research has been supported by funding from the United States National Science Foundation (NSF) via grants \#1942995 and \#1931258, as well as by the Department of Energy (DOE) Advanced Scientific Computing Research program via award DE-SC0023483.
Raghavendra Kanakagiri has been supported by NSF grant \#1931258 and the University of Illinois Urbana-Champaign Computer Science Future Faculty Fellows program.
This work used Stampede2 at TACC through allocation CCR180006 from the Advanced Cyberinfrastructure Coordination Ecosystem: Services \& Support (ACCESS) program, which is supported by National Science Foundation grants \#2138259, \#2138286, \#2138307, \#2137603, and \#2138296.
\end{acks}

\bibliographystyle{ACM-Reference-Format}
\bibliography{paper}



\end{document}